\documentclass[a4paper,fleqn,usenatbib,useAMS,usedcolumn]{mnras}

\usepackage{newtxtext} 

\usepackage[T1]{fontenc}
\usepackage{ae,aecompl}

\usepackage{graphicx} 
\DeclareGraphicsExtensions{.pdf,.png,.jpg,.ps}
\graphicspath{{./}}
\usepackage{amsfonts}
\usepackage{booktabs}
\usepackage{siunitx}
\usepackage{xcolor}
\usepackage{amssymb}  
\usepackage{hyperref}
\usepackage[nolist]{acronym}
\usepackage{ulem}
\usepackage{bm}
\usepackage{etoolbox}
\usepackage[inline]{enumitem}	
\usepackage{float}
\usepackage[linesnumbered,ruled]{algorithm2e}
\usepackage[]{algorithm2e}
\usepackage{amsmath}	

\makeatletter
\patchcmd\@combinedblfloats{\box\@outputbox}{\unvbox\@outputbox}{}{%
}%
\makeatother

\newlength\myindent
\usepackage{fontawesome} 

\usepackage{bbm}
\usepackage{mathtools}
\usepackage[export]{adjustbox}
\usepackage{pdfpages}
\usepackage[bottom]{footmisc}
\usepackage{hyperref}
\usepackage{graphicx}
\usepackage{pdflscape}

\definecolor{mustard}{rgb}{1.0, 0.86, 0.35}
\definecolor{cyan(process)}{rgb}{0.0, 0.72, 0.92}
\definecolor{ochre}{rgb}{0.8, 0.47, 0.13}
\definecolor{linesOne}{rgb}{0, 0.4470, 0.7410}
\definecolor{linesTwo}{rgb}{0.8500, 0.3250, 0.0980}
\definecolor{linesThree}{rgb}{0.9290, 0.6940, 0.1250}
\definecolor{linesFour}{rgb}{0.4940, 0.1840, 0.5560}
\definecolor{linesFive}{rgb}{0.4660, 0.6740, 0.1880}

\definecolor{deeppink}{HTML}{FF024F}
\definecolor{AISgray}{HTML}{696969}
\definecolor{AISblue}{HTML}{005AFF}
\definecolor{AISgreen}{HTML}{01FE01}
\definecolor{AISdarkgreen}{HTML}{4DAF4A}
\definecolor{AISpink}{HTML}{DC143C}

\definecolor{PopRed}{HTML}{FF6666}
\definecolor{PopGreen}{HTML}{65A165}
\definecolor{PopBlue1}{HTML}{87B1D3}
\definecolor{PopBlue2}{HTML}{66D8FF}

\definecolor{Verblue}{HTML}{377EB8}
\definecolor{Verorange}{HTML}{FF7F00}
\definecolor{Vergreen}{HTML}{4DAF4A}
\definecolor{Verpink}{HTML}{F781BF}

\definecolor{LightBlue}{HTML}{CBF1F1}

\definecolor{linkcolor}{rgb}{0.1216,0.4667,0.7059}

\newcommand\Fiducial{\texttt{Fiducial }}
\newcommand\rate{\mathcal{R}}

\newcommand{\AISs}{\textsc{STROOPWAFEL}}
\newcommand{\pluseq}{\mathrel{+}=}
\newcommand{\NEhits}{$N_{\text{T,expl}}$ }
\newcommand{\NE}{$N_{\text{expl}}$ }

\newcommand*\dif{\mathop{}\!\mathrm{d}}

\newcommand\fexplOne{$0.23$}
\newcommand\EffExplOne{$6.78  \cdot 10^{-3}$}
\newcommand\EffRefOne{$3.05   \cdot 10^{-1}$}
\newcommand\NxMCOne{$6.71  \cdot 10^{3}$}
\newcommand\NxAISOne{$2.35 \cdot 10^{5}$}
\newcommand\gainOne{$35 \times$}

\newcommand\fexplTwo{$0.27$}
\newcommand\EffExplTwo{$5.25 \cdot  10^{-3}$}
\newcommand\EffRefTwo{$3.69 \cdot  10^{-1}$}
\newcommand\NxMCTwo{$5.16  \cdot  10^3$}
\newcommand\NxAISTwo{$2.71 \cdot  10^5 $}
\newcommand\gainTwo{$53 \times$}

\newcommand\fexplThree{$0.66$}
\newcommand\EffExplThree{$6.36 \cdot  10^{-4}$}
\newcommand\EffRefThree{$7.38 \cdot 10^{-2}$}
\newcommand\NxMCThree{$6.55 \cdot  10^{2}$}
\newcommand\NxAISThree{$2.55 \cdot  10^{4}$}
\newcommand\gainThree{$39 \times$}

\newcommand\fexplFour{$0.59$}
\newcommand\EffExplFour{$9.03 \cdot  10^{-4}$}
\newcommand\EffRefFour{$9.71 \cdot 10^{-2}$}
\newcommand\NxMCFour{$8.93 \cdot  10^{2}$}
\newcommand\NxAISFour{$4.00 \cdot  10^{4}$}
\newcommand\gainFour{$45 \times$}

\newcommand\fexplFive{$0.69 $}
\newcommand\EffExplFive{$5.45 \cdot  10^{-4}$}
\newcommand\EffRefFive{$3.55 \cdot  10^{-1}$}
\newcommand\NxMCFive{$5.44 \cdot  10^{2}$}
\newcommand\NxAISFive{$1.10 \cdot  10^{5}$}
\newcommand\gainFive{$203 \times$}

\newcommand\fexplSix{$0.77 $}
\newcommand\EffExplSix{$3.43 \cdot  10^{-4}$}
\newcommand\EffRefSix{$3.38 \cdot  10^{-2}$}
\newcommand\NxMCSix{$3.32 \cdot  10^{2}$}
\newcommand\NxAISSix{$7.95 \cdot  10^{3}$}
\newcommand\gainSix{$24 \times$}

\newcommand{\CenterZero}{[20, 34, 0.3]}
\newcommand{\EpsilonZero}{[1.9, 8.0, 0.1]}
\newcommand{\CenterOne}{ [40, 1.0, 0.3]}
\newcommand{\EpsilonOne}{[1.7, 0.6, 0.2]}
\newcommand{\CenterTwo}{[34, 7.0, 0.8] }
\newcommand{\EpsilonTwo}{[1.8, 0.6, 0.1]}
\newcommand{\TrueFraction}{0.0013127}

\title[STROOPWAFEL: Simulating rare events]{STROOPWAFEL: Simulating rare outcomes from astrophysical populations,
 with application to gravitational-wave sources{\thanks{\AISs: Simulating The Rare Outcomes Of Populations With AIS For Efficient Learning.}}} %
\author[Broekgaarden et al.]
{Floor S. Broekgaarden,$^{1-5}${\thanks{E-mail: fbroekgaarden@g.harvard.edu} }
Stephen Justham,$^{7,8,1,2,5}$
Selma E. de Mink,$^{6,1,2}$
\newauthor
Jonathan Gair,$^{9,10,5}$
Ilya Mandel,$^{3,4,11,5}$
Simon Stevenson,$^{4,12}$
\newauthor
Jim W. Barrett,$^{13}$
Alejandro Vigna-G\'{o}mez,$^{11,3,4,5}$
Coenraad J. Neijssel$^{11,14}$
\\
$^{1}$Anton Pannekoek Institute for Astronomy, University of Amsterdam, Postbus 94249, 1090 GE Amsterdam, The Netherlands \\
$^{2}$GRAPPA, University of Amsterdam, Science Park 904, 1098 XH Amsterdam, The
Netherlands \\
$^{3}$Monash Centre for Astrophysics, School of Physics and Astronomy, Monash University, Clayton, Victoria 3800, Australia\\
$^{4}$The ARC Center of Excellence for Gravitational Wave Discovery -- OzGrav\\
$^{5}$Dark Cosmology Centre, Niels Bohr Institute, University of
Copenhagen, Juliane Maries Vej 30, DK-2100, K\o benhaven \o, Denmark \\
$^{6}$Harvard-Smithsonian Center for Astrophysics, 60 Garden  Street,  Cambridge, MA 02138 \\
$^{7}$School of Astronomy $\&$ Space Science, University of the Chinese Academy of Sciences, Beijing 100012, China\\
$^{8}$National Astronomical Observatories, Chinese Academy of Sciences, Beijing 100012, China\\
$^{9}$School of Mathematics, University of Edinburgh, The King`s Buildings, Peter
Guthrie Tait Road, Edinburgh, EH9 3FD, UK\\
$^{10}$Max Planck Institute for Gravitational Physics (Albert Einstein Institute), Am M\"{u}hlenberg 1, Potsdam-Golm 14476, Germany\\
$^{11}$Birmingham Institute for Gravitational Wave Astronomy and School of Physics and Astronomy, University of Birmingham, \\ 
$^{}$ Birmingham, B15 2TT, United Kingdom\\
$^{12}$Center for Astrophysics and Supercomputing, Swinburne University of Technology, Hawthorn VIC 3122, Australia,\\
$^{13}$Klarna Bank AB (publ). Sveav\"{a}gen 46, 111 34 Stockholm\\
$^{14}$Albert-Einstein-Institut, Max-Planck-Institut f\"{u}r Gravitationsphysik, D-30167 Hannover, Germany 
}

\date{\today}

\pubyear{2019}

\begin{document}
\normalem

\label{firstpage}
\pagerange{\pageref{firstpage}--\pageref{lastpage}}
\maketitle

\begin{abstract} 
Gravitational-wave observations of double compact object (DCO) mergers are providing new insights into the physics of massive stars and the evolution of binary systems. Making the most of expected near-future observations for understanding stellar physics will rely on comparisons with binary population synthesis models.  
However,  the vast majority of simulated binaries never produce DCOs, which makes calculating such populations computationally inefficient.
We present an importance sampling algorithm, \AISs, that improves the computational efficiency of population studies of rare events, by focusing the simulation around regions of the initial parameter space found to produce outputs of interest.
We implement the algorithm in the binary population synthesis code COMPAS, and compare the efficiency of our implementation to the standard method of Monte Carlo sampling from the birth probability distributions. 
\AISs \ finds $\sim$25--200 times more DCO mergers than the standard sampling method with the same simulation size, and so speeds up simulations by up to two orders of magnitude. Finding more DCO mergers automatically maps the parameter space with far higher resolution than when using the traditional sampling. 
This increase in efficiency also leads to a decrease of a factor $\sim$3--10  in statistical sampling uncertainty for the predictions from the simulations. 
This is particularly notable for the distribution functions of observable quantities such as the black hole and neutron star chirp mass distribution, including in the tails of the distribution functions where predictions using standard sampling can be dominated by sampling noise.  
\href{https://doi.org/10.5281/zenodo.3387651}{\color{linkcolor}\faBook}
\end{abstract}




\section{Introduction}
\label{sec:introduction}

The direct detection of gravitational waves  originating from merging binary black holes (BHs) and neutron stars (NSs) has opened up a new window on the Universe, and marked the birth of gravitational-wave astrophysics as a new field of research \citep[][]{PhysRevLett.116.061102, PhysRevLett.119.161101}. At the time of writing the first two observing runs of Advanced LIGO and Virgo have been completed \citep{2018arXiv181112907T}. A few dozen detections are expected during the third observing run, and we can anticipate hundreds of detections per year when the next generation of detectors with higher sensitivities come online \citep{2018LRR....21....3A}.

The detections are starting to reveal the properties of the population of merging binary BHs and NSs.  The distributions of the inferred masses and spins contain valuable information about their origin. 
Distinguishing different theories for their formation and learning about the complex physical processes that govern the lives of their possible massive-star progenitors requires comparing observed populations with theoretical predictions.  

Theoretical simulations of the population of merging DCOs are challenging because gravitational-wave events represent an extremely rare outcome of binary evolution.  
From a thousand massive binary systems typically only of order one, or less, yields a double compact object. A meaningful comparison with population observations requires simulating a statistically significant sample of events.  When sampling from the birth distributions, which is a form of sampling commonly used in binary population synthesis, this often means we need to sample at least many millions of binary systems. For example, \citet{kruckow2018progenitors} find  that  their DCO merger rates converge only when simulating $N \geq 3 \times 10^8$ binaries (and for BH--BH mergers their statistical noise remains at the 2 percent level even with $N=10^9$ samples).  

To make it feasible to simulate such large numbers of systems, all present-day simulations pay a high price.  In many studies computational speed is ensured by using highly approximate algorithms that treat the physical processes in a simplified way. 
Another way to keep the computational costs reasonable is to limit the total number of simulations, restricting the exploration of the impact of the uncertain physical input assumptions beyond a few variations. 

The recent detections bring to light a further challenge as we start to ask questions about rare subsets of the already rare gravitational-wave events.  One example is the subset of heavy binary black hole mergers with total system masses in excess of 50 $\rm M_{\odot}$.  Such systems produce loud GW signals and can thus be observed over large volumes \citep[][]{2017ApJ...851L..25F}. The majority of currently observed BH--BH mergers have total masses above 50 $\rm M_{\odot}$  \citep[][]{2018arXiv181112907T}. 
However, they are very rare in simulations of binaries sampled from the expected distribution of initial conditions.  Most current theoretical predictions for the extreme tails of the mass distribution are heavily affected by Poisson noise, resulting from under sampling.    A second example of an astrophysically important rare subset of the rare gravitational-wave events are the NS--NS systems that merge within about 50 $\rm Myrs$ from the moment the NS--NS is formed. 
Early NS--NS mergers are important as they are candidate sources for the observed early r-process enriched ultra-faint dwarf galaxies such as Reticulum II and Tucana III \citep[e.g.][]{2019ApJ...872..105S}.
 A third example  are the  subset of BH--NS mergers with sufficiently similar components masses that there is significant tidal ejection from the merger to  produce electromagnetic counterparts \citep{2018arXiv180700011F}.
Obtaining statistically accurate predictions for the extreme tails of the distribution functions for rare but astrophysically important subpopulations is currently a challenge for most simulations.   

Earlier studies have  proposed improvements of the efficiency of population synthesis studies.  \citet{1993A&A...271..149K} and later \citet{1996ApJ...465..338P} implemented a transformation function using Jacobian matrices to map known birth rates of cataclysmic variables directly into present day populations.  \citet{1996ApJ...471..352K} adopted this method, developed an analytical model for the kick prescription and showed that it is possible to obtain similar expressions for several observable distribution functions \citep{1998ApJ...493..351K,2000ApJ...541..319K}. More recently, \citet{andrews2017dart_board} implemented Markov Chain Monte Carlo (MCMC) methods to efficiently simulate populations of binaries matching specified evolutionary endpoints, whereas \citet{2017IAUS..325...46B} and \citet{2018arXiv180608365T} use Gaussian process emulators to predict outputs of the binary population synthesis model for parametrised choices of physical assumptions that have not been simulated.  
However,  current binary population synthesis models have complex output functions containing natural bifurcations (e.g.,  small changes  in the initial mass of a star can lead to drastic changes in the final mass in the simulations). Moreover, binary population synthesis simulation output spaces often contain stochastic behaviour (e.g., due to the randomly drawn neutron star natal kick). Such discontinuities pose a challenge for MCMC methods and Gaussian process regression emulators, as they rely on a certain smoothness in order to converge and produce independent samples.  

In this paper we present a new algorithm, \AISs$^{\ast,}$\footnote{All data that has been used for this study, accompanied with a Jupyter notebook that reproduces the main results, can be found here  \href{https://doi.org/10.5281/zenodo.3387651}{\color{linkcolor}\faBook}. The code for the \AISs{}  algorithm will be made publicly available after acceptance. Early inquiries can be addressed to the lead author.}.  
We have designed \AISs{} to improve the efficiency of simulations of rare astrophysical events, and so to enable accurate simulations of populations of extremely rare events at reasonable computational cost.
The algorithm first explores the initial parameter space until it finds a preliminary population of systems of interest. This exploration is done by stochastically sampling from the birth distributions.    \AISs \ then concentrates the later sampling towards regions in the initial parameter space that are in the vicinity of the initial parameters of the interesting binaries found during the exploration phase. This is an example of ``Adaptive Importance Sampling'' (AIS), described further in the next section.    

We focus here on the application of the study of DCO mergers as gravitational-wave sources, but our algorithm is much more broadly applicable.  The user can specify any target population of interest.  An advantage of the algorithm is  that it can handle the bifurcations and stochastic behaviour that naturally occur in the physical prescriptions in binary population synthesis simulations, and which lead to discontinuous output surfaces.   Finally, with our algorithm we can easily derive the uncertainties on the estimated parameters which can be a challenge for sampling methods such as MCMC and Gaussian Process regression emulators.

This paper is organized as follows.   In Section \ref{sec:sampling-methods} we describe the algorithm and provide expressions for how to calculate statistical estimates and the uncertainties from the simulations. We further derive the optimal relative duration of the exploratory phase and the total number of simulations, given the rareness of the target population.  In Section  \ref{sec:Results}  we provide a demonstration of our algorithm. We apply it to population synthesis simulations of double compact object mergers. 
In Section \ref{sec:discussion} we outline caveats and future directions for further improvement and refinement of the algorithm. We conclude and summarise in Section \ref{sec:conclusion}.

\section{Method}
\label{sec:sampling-methods}
Our algorithm is conceptually simple. It uses a strategy that may be familiar from playing the classic game \textit{Battleship}. The aim of this game is to guess the coordinates of the ships of the other player, which are placed on a regular discrete grid.  Most players will probably start with an exploration phase randomly trying different coordinates.  After one or more successful ``hits'' most players will change their search strategy and instead try to refine their search by trying coordinates that are close to the successful hits until they  uncover the full location and orientation of the ship. It has been shown that this is a more successful strategy compared to searching  completely randomly throughout the entire game   \citep[e.g.][]{jones1977blindfold}.

Our algorithm, \AISs, follows a conceptually similar strategy, but instead of aiming to win a game of Battleship the algorithm is designed to improve the efficiency for simulating populations of rare events (that is, rare outcomes from the space of initial conditions).  Successful hits in this analogy are finding systems of interest that are part of a certain target population. These may be systems that result in DCO mergers or anything that the user specifies. We improve the efficiency by focusing on areas of the initial parameter space near to those which produced outcomes of interest during a prior, exploratory, sampling phase.   Instead of Monte Carlo sampling from the birth probability distributions, \AISs{} uses information from that exploratory sampling phase to create an alternative distribution function, from which it then samples.

This class of Monte Carlo methods is generically called ``Importance Sampling" \citep{kahn1951estimation, 10.2307/166789}.  Since we do not know in advance which areas of the initial parameter space should receive extra attention we use \emph{adaptive} importance sampling (AIS), for which see, e.g., \citet{torrie1977nonphysical, doi:10.1080/00401706.1995.10484303, ortiz2000adaptive, pennanen2006adaptive, cappe2004population, cornuet2012adaptive}.  
The nature of the AIS algorithm makes it straightforward to tune the focus of the simulation on a specific target population or function of interest  \citep{2007arXiv0710.4242C}.
Such AIS algorithms also allow for straightforward calculations of the sampling uncertainties. 
 The \AISs \ implementation of AIS is similar to that in \citet{cornuet2012adaptive}, but includes a new method to guide the fraction of the computational effort that should be spent on the exploratory phase (see Sect~\ref{subsec:durationExploratory}). 

Whilst the concept of our algorithm is not complicated, there are some mathematical details involved in making the implementation efficient and robust.  For example, some of the subtlety involved 
is in making sure not to concentrate \emph{too} closely on locations which previously led to success. If we only look exactly where we have looked before, then we don't learn anything new.

Section \ref{sec:BPS} introduces binary population synthesis as a mapping between input and output parameter spaces, along with some notation which is useful for the description of our algorithm.  Section \ref{subsec:AISalgorithm} presents the key details of  \AISs .  We explain how we shape the adaptive sampling distribution from the information found in an initial exploratory phase, and how to optimally combine the samples from both the exploratory and adapted phases to estimate the population quantities of interest. We also describe how \AISs \ self-consistently determines how long the exploratory phase should last as a fraction of the simulation time, based on continually updated estimates of the rareness of the target population.
Section \ref{subsec:characteristics} illustrates the practical characteristics of our AIS algorithm in an idealised way, providing an explanatory summary of the behaviour of \AISs \ for users who do not wish to learn all the mathematical details.

\subsection{Definition of concepts and symbols}
\label{sec:BPS}
Binary population synthesis models the population observables for a particular class of event, under a set of assumptions about the physics.  Predicting such an output population typically involves simulating many individual systems from their initial conditions.  Only a small fraction of those simulated systems may produce outcomes which are of interest for that study.  

Selecting which specific points in the input space (i.e., the initial conditions) to simulate into the output space (i.e., the observables) is a key part of population synthesis.  This process is called sampling, and must appropriately take into account the relative frequency of different initial conditions.  Ideally, it should also efficiently explore the initial parameter space.  Examples of these initial parameters are the initial masses of the two stars, $m_{1,i}$ and $m_{2,i}$, and the initial separation $a_i$ between the two stars.  For a given initial composition, these three dimensions are often regarded as adequate initial conditions.  However, more generally, these input conditions may be distributed over many dimensions.   

Each initial binary system $\boldsymbol{x_i} \in \{ \boldsymbol{x_1}, ...,\boldsymbol{x_N} \} $ in a binary population synthesis can thus be written as $\boldsymbol{x_i} = (m_{1,i}, m_{2,i}, a_i, ...)$, which has a combined birth distribution 
\begin{equation}
	\pi(\boldsymbol{x_i}) = \pi({m_{1,i}}, m_{2,i}, {a_i},   \cdots).
	\label{eq:prior-3d} 
\end{equation} 
%
\begin{figure*}
	\includegraphics[width=0.9\textwidth]{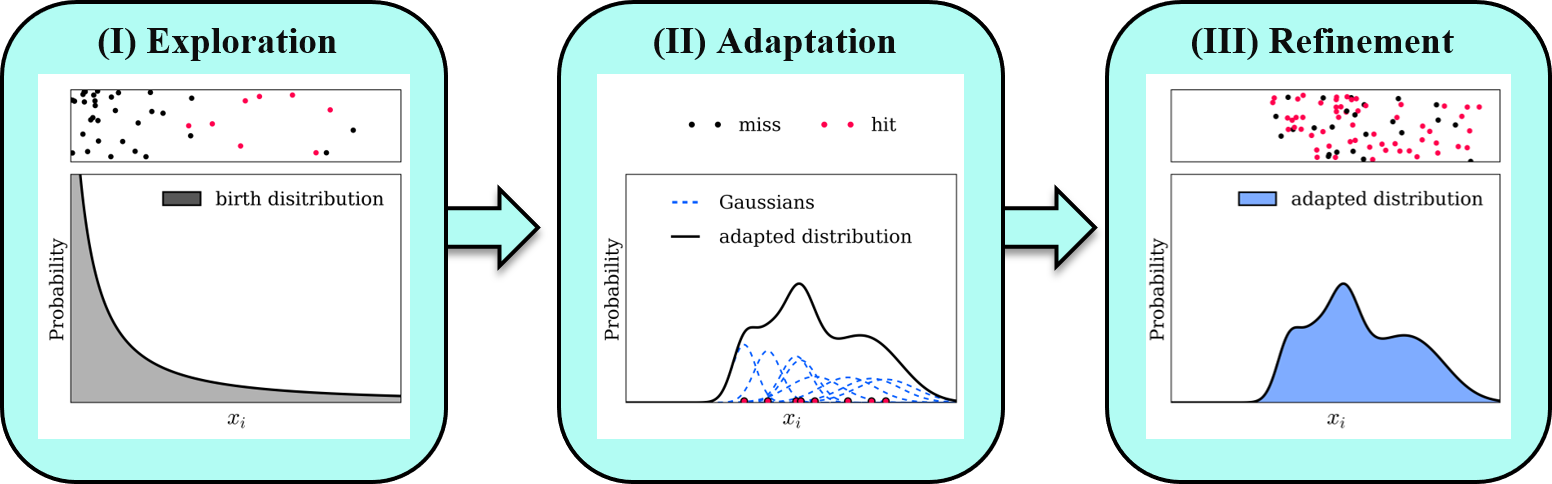}
    \caption{
    Illustration of the \AISs \ algorithm.  In the algorithm, (I)  we first draw random binaries from the birth distribution $\pi$  until a small population of events of interest (hits)  is found. (II)   We construct Gaussian distributions around each of the previously found successful events.  We create an adapted instrumental distribution $q(\boldsymbol{x})$   from  the mixture of Gaussian distributions. We scale the width of each Gaussian with the local sampling density. (III) We draw the remaining  samples from this adapted distribution which focuses  around the target population. The top panels show a random draw of samples from the corresponding distribution in the lower panel. The samples are assigned a random scatter in the y-direction for the visualization.}
    \label{fig:aIS_scheme}
\end{figure*}
This distribution of initial conditions is often taken as the Monte Carlo sampling distribution\footnote{Binary population synthesis simulations  often sample in the initial mass ratio,~$m_{2,i} / m_{1,i}$,  rather than the mass of the initially least-massive star. This includes the simulations we present later; see also Appendix ~\ref{app:COMPASdetails}. Whether sampling in $m_{2,i}$ or the mass ratio it has also been common to assume that the initial parameters are independent of each other, e.g.,:  $\pi(\boldsymbol{x_i}) = \pi({m_{1,i}}) \cdot$ $\pi({m_{2,i}} / m_{1,i})$ $\cdot  \pi({a_i}) \cdot \hdots $.  \newline This assumption of separability may not be valid, as found by \citet{1990ApJS...74..551A,2017ApJS..230...15M, 2018A&A...619A..77K}.}.   
In practice, simulations of binary-star populations which aim to study outcomes such as mergers between BHs and NSs do not sample from the full range of initial conditions of stellar binaries.  Such simulations ignore stars whose mass is too low to produce a BH or NS, which is a simple form of importance sampling.  The normalisation of $\pi$ actually used for the sampling is then corrected to take this into account when predicting event rates.

For each initial binary, $\boldsymbol{x_i}$, the final state of the binary  $\boldsymbol{y_f}$ is determined using the binary population synthesis model $u$,  
\begin{equation}
	\boldsymbol{y_f} = u(\boldsymbol{x_i}). 
	\label{eq:output-input} 
\end{equation}  
In many cases a simulation is run to study binaries  that evolve to a certain target subtype $\boldsymbol{T}$, e.g., maybe $\boldsymbol{T}$ is the population of binary black hole mergers.  The following indicator function describes whether a binary $\boldsymbol{y_f}$ is of interest:
\begin{equation}
\begin{aligned}
	\mathbbm{1}_{T}(\boldsymbol{y_f} ) \coloneqq \begin{cases} 1 \hspace{0.1cm} &\text{if } \boldsymbol{y_f} \in \boldsymbol{T}  \text{ (a hit) }  \\
	0  &\text{if } \boldsymbol{y_f} \notin \boldsymbol{T} \text{ (a miss)},  
	\end{cases}
	\label{eq:unityfunction}
\end{aligned}
\end{equation}
which equals 1 if $\boldsymbol{x_i}$ simulates to the target binary system
$\boldsymbol{T}$ (a hit) and zero if not (a miss).  Combining equations (\ref{eq:output-input}) \& (\ref{eq:unityfunction}) gives the function
\begin{equation}
\phi(\boldsymbol{x_i}) \coloneqq \mathbbm{1}_{T}(u(\boldsymbol{x_i}) ),
\end{equation} 
which is a shorthand notation to describe whether an initially drawn binary evolved into a binary of the target population.  

The samples from the initial parameter space that produced a binary of the target population (i.e.,  $\phi(\boldsymbol{x_i}  )= 1 $)  can then be given by the set
\begin{equation}
	\boldsymbol{x_{\text{T}}} \coloneqq ( \boldsymbol{m_{1,{\textbf{T}}}, \ m_{2,{\textbf{T}}}, \ {a_{\textbf{T}}}, \ \ldots }).
\end{equation}

At the end of a simulation, the properties of the model population and the statistical uncertainties on those predicted properties can both be determined 
using the standard Monte Carlo estimator  \citep{osti_4423221,doi:10.1080/01621459.1949.10483310}. For example, the relative formation rate of the target population, $\rate_{\text{T}}$, is estimated with
\begin{equation}
\mathbb{E}_{\pi}[\rate_{\text{T}}] \approx  \frac{1}{N_{\text{}}} \sum_{i=1}^{N_{\text{}}} \phi(\boldsymbol{x_i}),   
	\label{eq:unbiased}
\end{equation}
where $N$ is the total number of samples used, $\mathbb{E}$ is the notation for the estimated mean and the subscript $\pi$ in $\mathbb{E}_{\pi}$ denotes that the samples $\boldsymbol{x_i}$ are distributed following the birth distribution $\pi$ (cf. Eq. \ref{eq:prior-3d}).
We shall refer to this relative formation rate, $\rate_{\text{T}} $, throughout this section. Mathematically it is a fractional volume from the initial binary parameter space, weighted by the probability of forming a binary system at each part of that initial parameter space.  This quantity is not a physical rate, but gives a formation rate for the population of interest as a fraction of the total number of initial binary systems formed.  So it only differs from a true formation rate by a physical normalisation.  We consider it appropriately intuitive to keep referring to this as a fractional, or relative, rate.

\subsection{Adaptive sampling algorithm to increase efficiency of simulation}
\label{subsec:AISalgorithm}
Our algorithm consists of three main steps, as illustrated in Fig. \ref{fig:aIS_scheme}:
\begin{itemize}
	\item[]\textbf{(I)} \textbf{Exploration }\\  We first explore the parameter space by sampling directly from the birth distribution $\pi$ 
	 	 until eventually a sufficient population of events of interest is found. 
	\item[]\textbf{(II)} \textbf{Adaptation} \\ 
	 We construct multivariate Gaussian distributions in the initial parameter space around each of the events of interest found during the exploration phase.  We scale the widths of each of the Gaussians with the local sampling density. We create the adapted sampling distribution $q$, from here on referred to as the \emph{instrumental} distribution, by combining the Gaussians into a mixture distribution. 
	\item[]  \textbf{(III)} \textbf{Refinement} \\
	We draw the samples for the remaining simulations from this instrumental distribution. Each sample is assigned a weight so that the predicted population appropriately reflects the birth distribution $\pi$. 
\end{itemize}
The rest of this subsection explains these steps in more detail.

\subsubsection{The instrumental distribution}
\label{subsec:instrumental-distribution}
When the exploratory phase has ended, the set of binaries of the target population $\boldsymbol{x_{\rm{T}}}$  (i.e., hits) contains \NEhits  binaries that were found using a total of \NE samples. In the \AISs \ algorithm these samples are then used to create an adapted instrumental distribution $q(\boldsymbol{x})$, which is focused around the areas in the initial parameter space that produced the binaries of interest during the exploratory phase.  The remaining binaries are thereafter sampled from this instrumental distribution.  To obtain unbiased estimates of the target population, weights ${w_i}$ are incorporated for each sample as is standard in importance sampling
\begin{equation}
	{w_i} = \frac{\pi(\boldsymbol{x_i})}{q(\boldsymbol{x_i})},
	\label{eq:weights}
\end{equation} 
where $\pi$ is the distribution of initial conditions, as given in Eq. (\ref{eq:prior-3d}). 

The instrumental distribution $q(\boldsymbol{x})$ can be chosen to be any probability distribution function, but a robust instrumental distribution is characterized by the following criteria:
\begin{itemize}
	\item The weights $w_i$ are always finite and well defined. That is, $q(\boldsymbol{x_i}) = 0$ implies  $ \pi(\boldsymbol{x_i})\phi(\boldsymbol{x_i})  =0$ for all $i$.
	\item The instrumental distribution is efficient if  $q(\boldsymbol{x})$ is close to the (unknown) target distribution of the binary population synthesis study, i.e., when the instrumental distribution $q(\boldsymbol{x}) $ is proportional to $ | \phi(\boldsymbol{x}) | \pi(\boldsymbol{x}) $, as shown by \citet{10.2307/166789}. 
	\item It should be computationally inexpensive to generate random samples from $q(\boldsymbol{x})$ as well as to calculate the probability $q(\boldsymbol{x_i})$ for each sample $\boldsymbol{x_i}$. 
\end{itemize}

In order to achieve these properties the instrumental distribution, $q(\boldsymbol{x})$, in \AISs \ is chosen to be a mixture\footnote{In this context, ``mixture" has a standard mathematical meaning. A sample drawn from $q$ is drawn from each Gaussian $q_k$ with a probability $N_{\text{T,expl}}$ instead of taking the sum of normally distributed samples. The sum of two jointly normally distributed random variables will still have a normal distribution (even if the means are not the same) whereas a \emph{mixture} of two normally distributed variables will have two peaks (assuming the means are far enough apart).} of \NEhits Gaussian distributions $q_k$ 
given by
\begin{equation}
	q(\boldsymbol{x}) =  \frac{1}{ \text{\NEhits}} \sum_{k={1}}^{\text{\NEhits}}  q_k({\boldsymbol{x}; \boldsymbol {\mu _{k},\Sigma _{k}}}),
	\label{eq:instrumental}
\end{equation}
where each $q_k$ contributes $1 / \text{\NEhits}$ to the mixture distribution. 

However, when drawing from $q(\boldsymbol{x})$, some samples will fall outside the physical range of the parameter space  $\Omega$ (e.g., when drawing a binary with a negative stellar mass). Such samples  can immediately be rejected and redrawn. By doing so, we in practice sample from the normalized physical mixture distribution
\begin{equation}
	\widetilde{q(\boldsymbol{x})} =  \frac{1}{(1-F_{\rm{rej}})} \, q(\boldsymbol{x}) \, \mathbbm{1}_{\Omega}(\boldsymbol{x} ) 
	\label{eq:instrumental}
\end{equation}
where $\mathbbm{1}_{\Omega}(\boldsymbol{x})$ is the indicator function that equals 1 when the sample $\boldsymbol{x}$ lies in the physical range of the parameter space and 0 if not. The factor $ F_{\rm{rej}}$ is the fraction of samples from $q(x)$ that are drawn outside of the physical parameter space.  The factor $1/ (1 - F_{\rm{rej}})$  thus corrects for the normalization of $\widetilde{q(x)}$.     It is computationally inexpensive to draw samples from $\widetilde{q(x)}$ since $F_{\rm{rej}}$ can be estimated once with a Monte Carlo simulation, and one can draw randomly from each Gaussian $q_k(x)$ separately. 

The Gaussian distributions  $q_k({\boldsymbol{x}; \boldsymbol {\mu _{k},\Sigma _{k}}})$ are parametrized by their means $\boldsymbol{\mu_k}$ and covariance matrices $\boldsymbol{\Sigma_k}$. 
The covariance matrix, $\boldsymbol{\Sigma_k}$, determines the width of the Gaussian distributions. We adopt a diagonal covariance matrix  
\begin{equation}
	\boldsymbol{\Sigma_k} = \begin{bmatrix} 
    		\sigma_{1,k}^2 	& 0 	& \dots \\
    		0 				& \ddots & \\

    		\vdots	 			&      & \sigma_{d,k}^2 
    \end{bmatrix}, 
	\label{eq:covariance-matrix}
\end{equation}
where $d$ is the number of dimensions of the initial parameter space.  

We scale the width of each Gaussian, given by the  covariance matrix $\boldsymbol{ \Sigma_k}$,  with the average distance to the next sampled binary $ \boldsymbol{x_i}$ in the initial parameter space, estimated via the local density of the prior distribution $\pi$. This allows the algorithm to construct broader Gaussian distributions (i.e., with larger $\sigma_k$) around previously found hits that lie in the regions of the parameter space that are less densely explored.  
If the one dimensional marginalised prior of the $j$-th parameter is $\pi_j$ then the standard deviation $\sigma_{j,k}$ is given by:
\begin{equation}
		\sigma_{j,k} = \kappa \frac{1}{\pi_j(x_k) N_{\rm{expl}}^{1/d}},    
\label{eq:sigmajk}
\end{equation}
where $N_{\rm{expl}}$ represents the number of samples used for the exploration phase,  the power ${1/d}$ scales this number to the effective number of samples per dimension and the factor $\pi_j(x_k)$ scales the width to the density of samples in the exploration phase around the previously found hit $x_k$. We also introduce a free parameter, $\kappa$, that scales the widths of the Gaussian distributions.  This enables us to regulate how tightly the mixture of Gaussian distributions covers the parameter space near the successful binaries $\boldsymbol{ x_{\text{T}}}$.
In this paper we adopt $\kappa = 2$, which we chose following tests with a toy model (for which, see Section \ref{subsec:kappaAndToymodel} and Appendix~\ref{app:toymodel}).

\subsubsection{Combining samples from the exploratory phase and refinement phase}
\label{subsec:CombiningSamples}
It is desirable to make use of the samples from both the exploration and refinement sampling phases.  The optimal way to achieve this is somewhat subtle. 
In principle this could be done by merging the samples and weights into a combined estimate \citep[see Chapter 14][]{robert2013monte}. However, \citet{Veach:1995:OCS:218380.218498},  \citet{doi:10.1080/00401706.1995.10484303}, and later \citet{doi:10.1080/01621459.2000.10473909} showed that using 
\emph{deterministic multiple mixture weights} is an efficient and robust way of combining the samples. This approach uses the fact that the combined samples from the exploratory phase and refined sampling phase can be represented by a mixture sampling distribution $Q(x)$ from both phases
\begin{equation}
	Q(\boldsymbol{x}) =  f_{\text{expl}}  \pi(\boldsymbol{x})  +  (1 - f_{\text{expl}}) \widetilde{q(\boldsymbol{x})}, 
\label{eq:CombinedDistribution}
\end{equation}
where $f_{\rm{expl}} = N_{\rm{expl}} / N $ is the fraction of samples spent on the exploratory phase.  %
By analogy with Eq.~(\ref{eq:weights}), the weights of all the $N$ samples can be recalculated with 
\begin{equation}
	\widetilde{w_i} = \frac{\pi(\boldsymbol{x_i})}{Q(\boldsymbol{x_i})}.
	\label{eq:weightsMixture}
\end{equation} 
This use of deterministic multiple mixture weights is not fundamental to \AISs{}.  Our motivation for using deterministic multiple mixture weights is conservative, to increase the stability against potential sampling artefacts.  One of the samples drawn from the importance function $q(x)$ may occasionally be extremely large. Such extreme weights could remain so large as to be problematic when merging the samples with the original weights $w_i$ -- no matter how efficient the sampling is in the exploratory phase  \citep{cornuet2012adaptive}.  Use of deterministic multiple mixture weights suppresses this potential rare difficulty.

The multiple mixture weights approach ignores the distribution from which a given draw was sampled. This does not affect the estimators for the predicted values, although it does introduce a very small bias to the uncertainty estimators, which we confirmed to be negligible in our toy model tests.  Recalculating the weights in this way yields comparable or better estimates than those which are obtained when merging the samples or using inverse-variance weighting for our adaptive importance sampling algorithm. Indeed, \citet{2014arXiv1411.3954H} derived a bound for the variance of the \textit{balance heuristic} for such estimators that combine samples from different distributions and found that this is an efficient way of combining samples.
 See also sect. 3 in \citet{Veach:1995:OCS:218380.218498} and sect. 2 in \citet{cornuet2012adaptive} for a more detailed discussion. 

\begin{table}
\centering
\begin{tabular}{|l|l|} \hline 
\textbf{symbol} & \textbf{description}    \\ \hline 
$u$ & {binary population synthesis model} \\ \hline
$\boldsymbol{x_i}$ & {initial state of a binary system} \\ \hline
$\boldsymbol{y_f}$ & {final state of a binary system} \\ \hline
$\boldsymbol{T}$ & {target subpopulation of binaries of interest} \\ \hline
$\boldsymbol{{x_{\rm{T}}}}$ & {set of hits from the exploratory phase} \\ \hline
$\pi(x)$ & {distribution of the initial conditions} \\ \hline
 ${\Omega}$ & {physical parameter range of the simulation} \\ \hline
$q(x)$ & {instrumental distribution } \\ \hline
$\widetilde{q(x)}$ & {normalized instrumental distribution } \\ \hline
$N$ & {total number of samples in the simulation} \\ \hline
$N_{\text{expl}}$ & {number of binaries in the exploratory phase} \\ \hline
$N_{\text{ref}}$ & {number of binaries in the refinement phase} \\ \hline
$N_{\text{T}}$ & {number of systems of the target population} \\ \hline
$N_{\text{T,expl}}$ & {number of hits in the exploratory phase} \\ \hline
$f_{\rm{expl}}$ & {fraction of samples in the exploration phase} \\ \hline
$\kappa$ & {scale factor for the widths of the Gaussians} \\ \hline
$w_i$ & {statistical weight of sample $\boldsymbol{x_i}$} \\ \hline
$\widetilde{w_i}$ & {recomputed statistical weight} \\ \hline  
\end{tabular}
\caption{Summary of the parameters that are used throughout this paper. Hits refer to binaries of the chosen target population.} 
\label{tab:variables}
\end{table}

\subsubsection{Calculating statistical estimates using the adaptive distribution}
\label{subsec:AISstatistic}
At the end of each run the properties of the target population, such as the rate of formation $\rate_{\text{T}}$ of members of the target population, and distribution functions of population observables, can be determined by standard Monte Carlo estimates.  Because we have drawn the samples from a different distribution than the birth distribution we have to incorporate weights to make sure that the estimators for these quantities reflect the correct formation probabilities.   
For example, the relative formation rate $\rate_{\text{T}}$ of the target population within the simulation is estimated by the mean
\begin{equation}
	\rate_{\text{T}} \approx 	{\mathbb{E}_Q[\rate_{\text{T}}]} = \frac{1}{N} \sum_{i=1}^{N} 
	\phi(\boldsymbol{x_i}) \widetilde{w_i}.	
	\label{eq:ISestimator}
\end{equation}
The uncertainty in this rate is represented by the variance about the mean by
\begin{align}
	{\mathbb{V}_Q[\rate_{\text{T}}]}
	= \frac{s^2_Q[\phi({\boldsymbol{x}})\widetilde{{w(\boldsymbol{x})}}] }{N} \approx \frac{\sum_i^N  \phi(\boldsymbol{x_i})^2  \widetilde{w_i}^2}{N} - \rate_{\text{T}}^2.
	\label{eq:variance}
\end{align}
where $s_Q$ is the (sample) standard deviation for samples drawn from the mixture sampling distribution $Q(\boldsymbol{x})$. 
This equation is known as the asymptotic variance.  
The other moments or statistical estimates for the target population can be similarly calculated.

\subsubsection{The duration of the exploratory phase}
\label{subsec:durationExploratory}
An important choice in adaptive importance sampling algorithms is deciding when to switch from the exploratory phase to the refinement phase. This choice can have a substantial impact on the performance of the algorithm. 
Leaving the exploratory phase too early can result in missing important regions of the initial parameter space which produce systems in the target population.   On the other hand, switching to the refinement sampling phase too late will miss out on the advantages of the algorithm, as most time will be spent sampling from the birth distributions instead of the more efficient adapted distribution. 

A method often used in adaptive importance sampling algorithms to determine the fraction of samples that should be spent on exploring the parameter space is by using the effective sample size: $(\sum_k {w_k})^2 / \sum_k({w_k}^2)$, \citep[ESS,][]{doi:10.1080/00401706.1995.10484303,liu2008monte}. This is a measure of efficiency and corresponds to the  equivalent number of independent  samples drawn from the prior distribution.
However, it can be difficult to know in advance what a good value for the ESS should be. Instead, since \AISs{} is a two-step adaptive algorithm, we can directly derive a value for $f_{\rm{expl}}$ by using the estimated rareness of the population, which we self-consistently calculate during the exploration phase. 

Here we estimate the optimal fraction $f_{\rm{expl}}$ of the total number of samples $N$ that we should spend on the exploratory phase.  The challenge is that we do not know in advance how good our adaptive sampling distribution will be.  Here, as a simplified proxy for the imperfect sampling distribution, we assume that the adaptive sampling distribution is determined sufficiently well that it perfectly matches the target distribution over most of the parameter space, but that a small region of the target parameter space could be missing samples due to a limited number of samples drawn during the exploratory phase, and thus have an adaptive sampling probability of zero (see Appendix \ref{app:Fprior} for details).  In other words, we divide the volume of the input parameter space which successfully produces systems of interest into two parts, one which we assume we have accurately found and one which remains missing. We then find $f_{\rm{expl}}$ such that the event rate uncertainty is minimised; specifically, we require that the contribution to the event rate from potentially undiscovered islands is smaller than, or no worse than similar to, the sampling uncertainty in the rate contributed by the islands which are successfully found.

We assume that we sample from the mixture distribution $Q(x)$, and aim to estimate the rate $\rate_{\text{T}} $ with $N$ total samples.   After simulating $N$ samples we assume we have identified a target binary forming region with total weight $z_1$, whereas a region with weight $z_2$ is yet undiscovered such that the estimated rate of the target population is 
\begin{equation}
	\rate_{\text{T}} =  \underbrace{z_1}_{\text{identified}} + \overbrace{z_2}^{\text{unidentified}} \approx {\mathbb{E}_Q[\rate_{\text{T}}]}. 	 
	\label{eq:EstimateFexpl}
\end{equation}
The uncertainty on this rate estimate is described by the variance, which we can approximate with (see Appendix \ref{app:Fprior} for more details)
\begin{equation}
	\mathbb{V}_Q[\rate_{\text{T}}] \approx \frac{1}{N} \left[  \frac{z_1}{\frac{(1-f_{\rm{expl}})}{z_1} + f_{\rm{expl}}}  + \frac{z_2}{f_{\rm{expl}}} - (z_1 + z_2)^2    		\right].	
	\label{eq:VarFprior}
\end{equation}

The smallest uncertainty on the rate $\rate_{\text{T}}$ is obtained for the value of $f_{\rm{expl}}$ where the variance $\mathbb{V}_Q(\rate_{\text{T}})$ is the lowest. 
By taking the derivative of $\mathbb{V}_Q$ with respect to $f_{\rm{expl}}$, and then finding the roots of the derivative, we find that this minimum occurs at
\begin{equation}
	f_{\rm{expl}} =  1 - \frac{z_1 (\sqrt{1-z_1} - \sqrt{z_2})}{\sqrt{1-z_1}(\sqrt{z_2(1-z_1)} +z_1)} . 
	\label{eq:FexplSolutionComplicated}
\end{equation}

To make practical use of this during our simulations, we need ongoing live estimates for $z_1$ and $z_2$.   For the region which has been identified, we adopt $z_1 \approx {\mathbb{E}_{\pi}[\rate_{\text{T}}]}$ during the simulation using Eq.~(\ref{eq:ISestimator}). We approximate the target population region that is yet undiscovered, $z_2$, by $z_2 \approx \frac{1}{(f_{\rm{expl}} N)}$. This represents the weight of stochastic sampling noise in the exploratory phase when using a total of $f_{\rm{expl}} N$ samples. 
Moreover, this choice of $f_{\rm{expl}}$ ensures that the estimated uncertainty on the rate estimate  is always comparable or larger than the uncertainty of missing a region, $ 1 / f_{\rm{expl}} N $. 

While the running estimate of $z_2$ on the right hand side of Eq.~\ref{eq:FexplSolutionComplicated}  is a function of $f_{\rm{expl}}$, rather than explicitly solving for  $f_{\rm{expl}}$, we choose to iteratively approach the optimal solution over the course of the exploratory phase. The resulting $f_{\rm{expl}}$ is similar to the $f_{\rm{expl}}$ that is obtained  if we had used the dependency of $z_2$ on  $f_{\rm{expl}}$ when solving for the minimum in Eq.~\ref{eq:VarFprior}. 

We note that we have implicitly assumed that the adaptive sampling phase is perfectly efficient, i.e., that all the draws in the adaptive sampling phase find a member of the target output population.   Here that is a conservative assumption, since a less-than-perfect efficiency will increase the sampling uncertainty with respect to the known islands (i.e., $z_1$).   Therefore, imperfect efficiency in the adaptive sampling phase decreases the chance that the uncertainty from undiscovered islands is significant.  

\AISs{} internally uses Eq.~\ref{eq:FexplSolutionComplicated} to estimate $f_{\rm{expl}}$.  For clarity here, we can additionally assume that $z_1$ and $z_2$ are much smaller than 1, i.e., that the target population is a rare outcome of the initial conditions.  Then we obtain the simplified equation
\begin{equation}
	f_{\rm{expl}}   \approx 1-  \frac{z_1}{z_1 + \sqrt{z_2}}    .
	\label{eq:FexplSolutionSimple}
\end{equation}

From this simplified equation it becomes clear that we recover the intuitively correct limit for extremely rare events, i.e. if  $z_1 = {\mathbb{E}_Q[\rate_{\text{T}}]} \rightarrow 0  $, we find $f_{\rm{expl}} \rightarrow 1$, which suggests that we should spend all our simulation time on exploration, as expected.  On the other hand once we find at least 1 hit in the exploratory phase we find $f_{\rm{expl}} \neq 1$, and so the variance of our rate estimate is expected to decrease when drawing some of the samples from the adapted distribution, compared to taking all samples from the birth distribution.  

Lastly, from Eq.~\ref{eq:VarFprior} it also becomes clear that $f_{\rm{expl}} \approx 0.5$ once we have found $N_{\rm{T,expl}} \sim \sqrt{N_{\rm{expl}}}$ target binaries. 
 In other words, $f_{\rm{expl}} \sim 1 $ if $N \leq 1/\rate_{\text{T}}^2$ and therefore the total number of samples $N$ should generally be similar to or larger than  $ 1/\rate_{\text{T}}^2$.

\subsubsection{Determining the free parameter $\kappa$ from tests with a toy model }
\label{subsec:kappaAndToymodel}
We present results from the application of our method to astrophysical simulations in Section \ref{sec:Results}. Here we explore the methodology with a toy model to test  the performance of the algorithm and determine the value of the free parameter $\kappa$.  In principle $\kappa = 1$ could be adopted. Smaller values of $\kappa$ will increase the efficiency of the \AISs{} algorithm, but increase the chance of missing an important region of the output surface because the Gaussian distributions $q_k$ are too narrow and do not cover the output surface well. Excessively large values of $\kappa$, meanwhile, will decrease the efficiency of finding samples of interest in the refinement phase and lower the gain of \AISs.  After performing tests with a toy model, as described in Appendix \ref{app:toymodel}, we adopt the value $\kappa =2$.However, when applying \AISs{} in higher dimensions the optimal value for $\kappa$ may well change.

\subsubsection{Summary of \AISs \ algorithm}
\label{subsubsec:algorithmflow}
The algorithm for \AISs{}, combining the methods discussed in this section, is summarized in Algorithm~\ref{alg:expl}.

\begin{algorithm}
    $i = 0 $\;
    	\texttt{\\}
	\textbf{(I) Exploration:}\\ 
	$f_{\rm{expl}} = 1$\;
    \While{ $i / N \leq f_{\rm{expl}} $}   
      {
		$i \pluseq 1$\; 
        draw new sample $x_i \sim \pi(x)$\;
        evaluate sample $y_f = u(x_i)$\;
        \If{$y_f \in T$}{
        counthits $\pluseq1$\;
        $\boldsymbol{x_{\text{T}}} \leftarrow \boldsymbol{x_i}$ (add hit to the found collection of hits)\; 
        update estimate $f_{\rm{expl}} $  iteratively using  Eq. 18 $z_1 = {\mathbb{E}_{\pi}[\rate_{\text{T}}]}$ and  $z_2 = \frac{1}{(f_{\rm{expl}} N)}$\;
        }         
      }
      \texttt{\\}
      \textbf{(II) Adaptation:}\\ 
      	set $\boldsymbol{\mu} = \boldsymbol{x_{\text{T}}}$\;
      	Calculate $\boldsymbol{\Sigma}$ by determining $\sigma_{j,k}(\boldsymbol{x_{\text{k}}})$ for all $k = 			{1}, ... ,  N_{{\text{T}}}$\; 
      	This gives $q(\boldsymbol{x}; \boldsymbol{\mu},\boldsymbol{\Sigma})$\;
	\texttt{\\}
      \textbf{(III) Refinement:}\\ 
    	\While{$  N_{\rm{expl}} \leq$  i   $\leq  N$ } 
      	{
			$i \pluseq 1$\; 
        	draw new sample $x_i \sim q(\boldsymbol{x}; \boldsymbol{\mu},\boldsymbol{\Sigma}) \mathbbm{1}_{\Omega}(\boldsymbol{x})$\;
        	evaluate sample $y_f = u(x_i)$\;
      	}    
            \texttt{\\}
      \textbf{Post  processing:}\\   
	calculate $F_{\rm{rej}}$ and mixture weights $\widetilde{w_i} = \frac{\pi(\boldsymbol{x_i})}{Q(\boldsymbol{x_i})} $\;
	calculate desired population quantities such as the rate $\rate_{\text{T}} $\;      
    \caption{\AISs \ algorithm}
    \label{alg:expl}
\end{algorithm}
%

\subsection{Characteristic behaviour}
\label{subsec:characteristics}

Here we use the analytic derivations from Section \ref{subsec:AISalgorithm} to illustrate the characteristics of our algorithm in idealised cases. This is intended to help users of \AISs \ understand the expected behaviour without needing to master the details presented above.  

Key variables for this illustration are:
\begin{itemize}
\item{$\rate_{\text{T}}$, the formation rate of the population under study. This is expressed as the fraction of binaries, when drawn from initial conditions following the birth probability distribution, that yield target systems.}
\item{$N$, the total number of binary systems (i.e. samples) used in a given population simulation, which is chosen by the user.}     
\item{$f_{\rm{expl}}$, the fraction of the total number of samples that should optimally be spent on the exploration phase. This is chosen automatically by \AISs \ during the exploration phase (see Sect. \ref{subsec:durationExploratory}),  when the algorithm estimates the formation rate $\rate_{\text{T}}$.}
\end{itemize}

\begin{figure}
	\includegraphics[width=1\columnwidth]{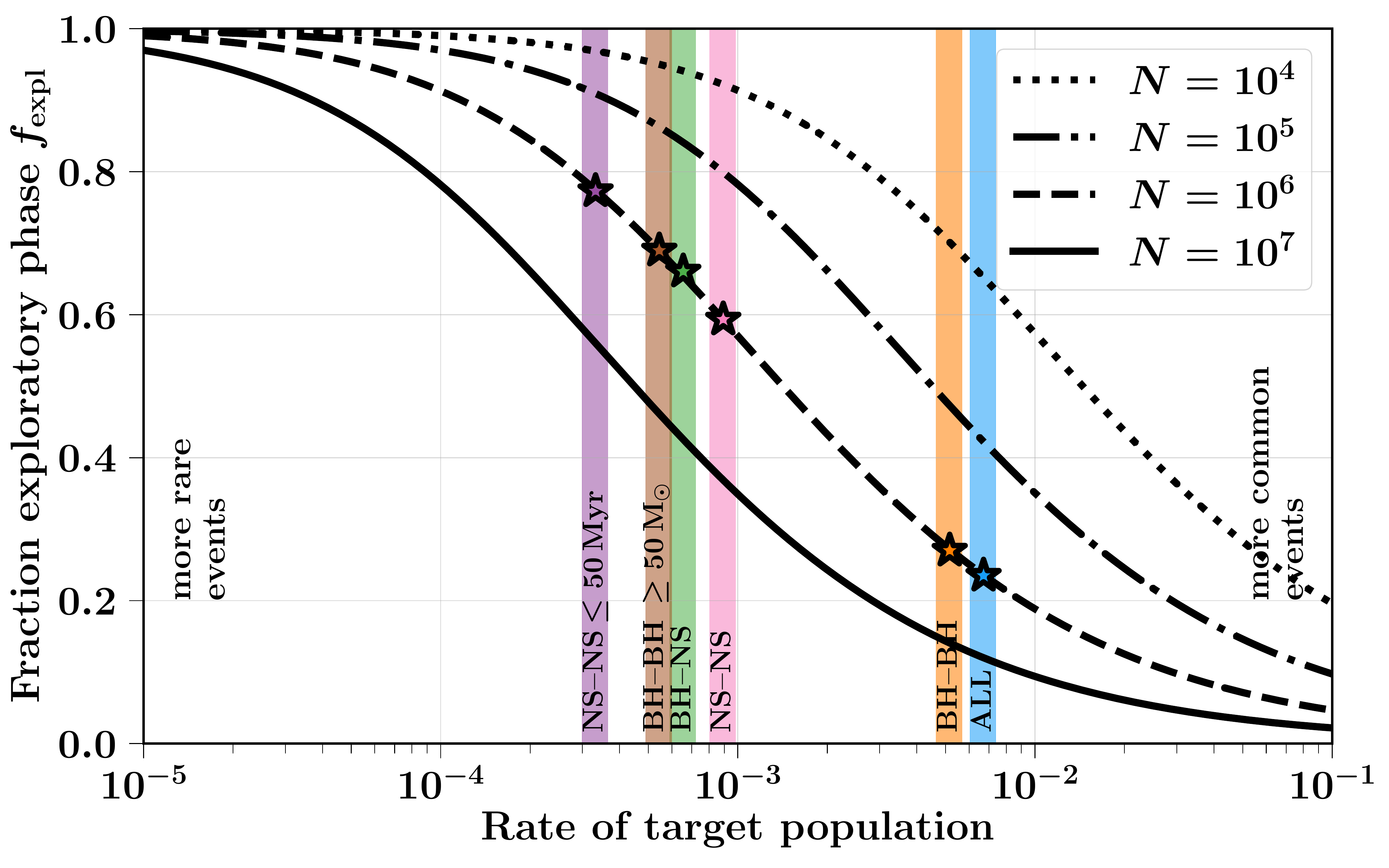}
	\includegraphics[width=1\columnwidth]{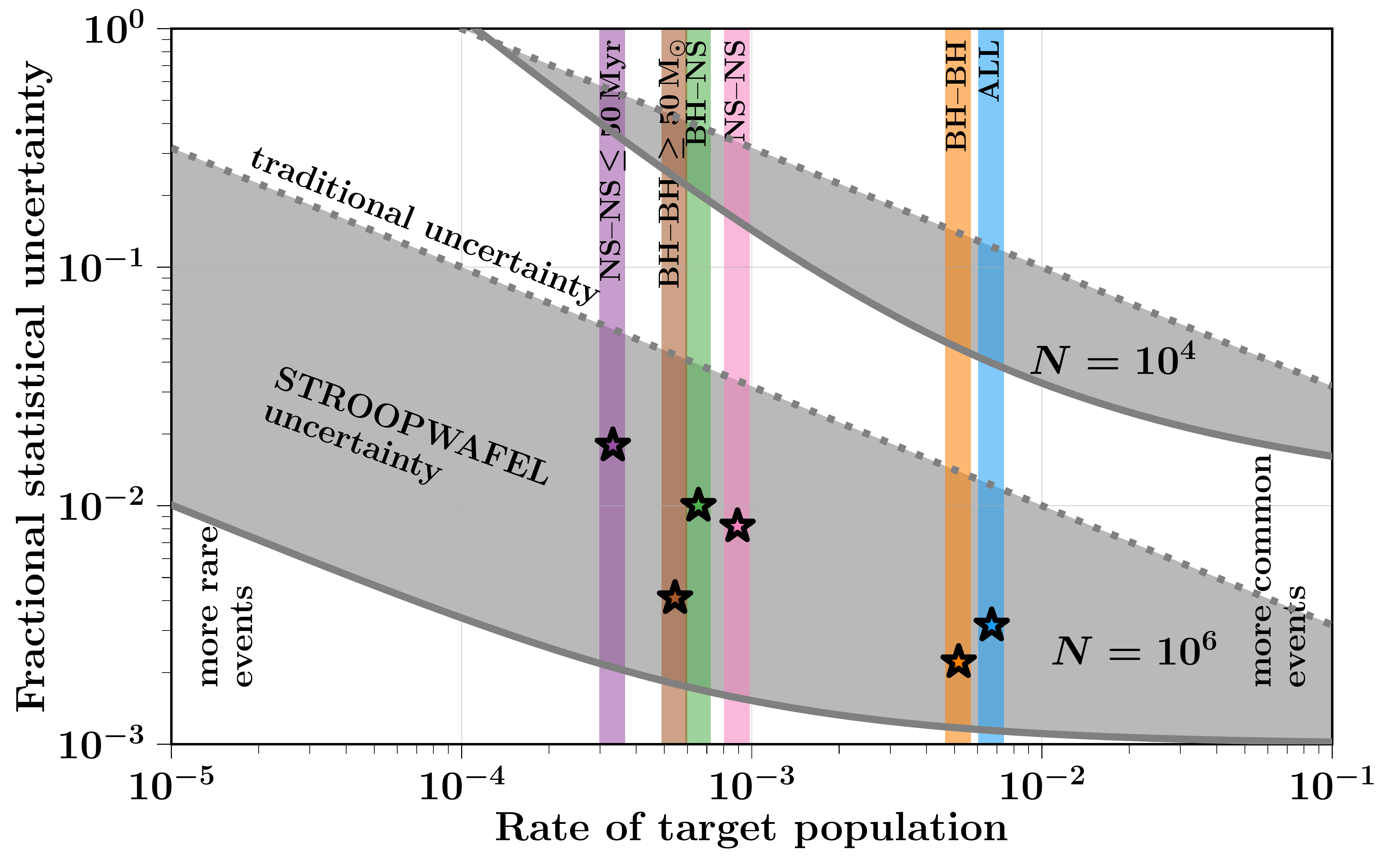}%
	    \caption{  	    
	    \textbf{Top panel:} the fraction of total samples that should be spent on the exploratory phase versus the fractional rate, $\rate_{\text{T}}$, of the target population in a simulation. Different curves show the effect of varying the total number of samples $N$ in the simulation. 
	    \textbf{Bottom panel:} expected sampling uncertainty on the predicted event rate versus the fractional rate, for two different choices of $N$. The dashed lines show the expected uncertainty from "traditional" Monte Carlo sampling (i.e., $\sqrt{{\mathbb{V}_{\pi}[\rate_{\text{T}}]}}/\rate_{\text{T}}$ ).  The solid lines show the minimum possible statistical uncertainty from \AISs{} sampling, i.e., if the efficiency in the refinement phase is 1.	 In practice the statistical rate uncertainty when using \AISs{} will lie in the area shaded in grey.  
	    \textbf{All panels:} coloured vertical bars indicate the fractional rate of the target populations, and star symbols show the corresponding values of these parameters, for the six simulations with $N= 10^6$ described in Section \ref{sec:Results}.}
    \label{fig:fractionPrior}
\end{figure}

\begin{figure}
	\includegraphics[width=1\columnwidth]{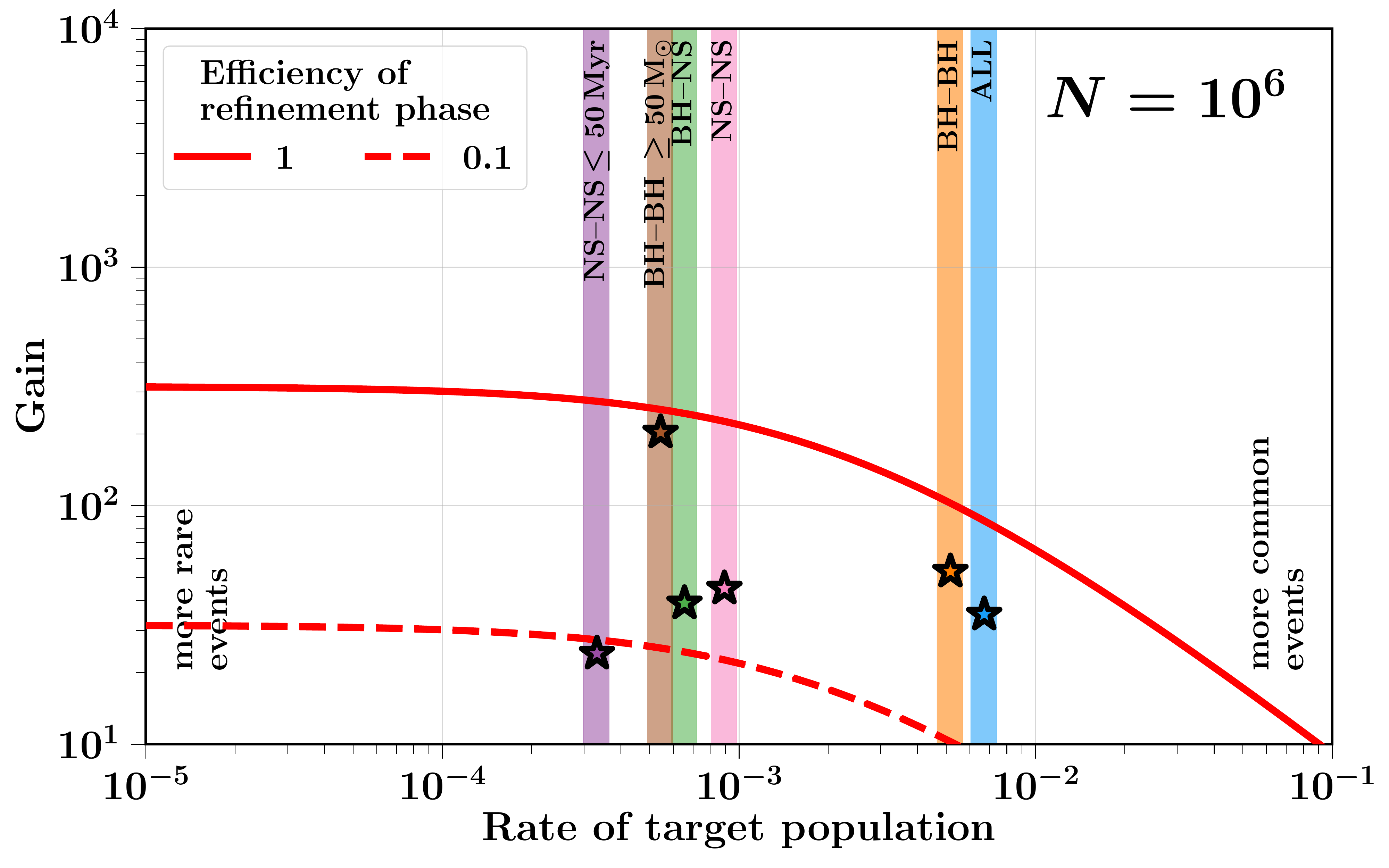}
	    \caption{The ratio of the number of target binaries found with \AISs{} to the number found when Monte Carlo sampling from the birth distribution -- i.e., the multiplicative gain achieved by \AISs{} -- as a function of rareness of the target population. The red curves give this gain for two different efficiencies in the refinement phase, as labelled. The gain is shown for simulations with a total of $N=10^6$ samples. Coloured vertical bars and star symbols present the values for the six simulations described in Section \ref{sec:Results}, as given in the last column of Table~\ref{tab:comparison2}. }
    \label{fig:gain-for-imperfect-efficiency}
\end{figure}

Figure~\ref{fig:fractionPrior} presents derived quantities as a function of the fractional rate of the target population $\rate_{\text{T}}$.  This Figure, and Figure~\ref{fig:gain-for-imperfect-efficiency}, include points representing simulated astrophysical populations, specifically for different subsets of DCO mergers. These simulations are described further in Section \ref{sec:Results}. The values from those simulations are included here to give context to the analytic expectations for the performance of \AISs{}.  

%

The top panel of Figure~\ref{fig:fractionPrior} shows the optimal $f_{\rm{expl}}$, for simulations with a total number of samples of $N = 10^4, 10^5, 10^6 $ and $10^7$.   For a rarer target population, a larger fraction of the total number of samples should be spent on the exploratory phase.  This is because it takes longer to determine a good sampling distribution when $\rate_{\text{T}}$ is low.
 A more common target population can be optimally simulated with a relatively small exploratory phase, since we expect this will be enough to build up a good adaptive distribution.    \AISs{} estimates $\rate_{\text{T}}$ during the exploration phase, when it samples from the birth probability distribution and self-consistently calculates $f_{\rm{expl}}$ based on the estimated $\rate_{\text{T}}$ and user-chosen $N$. 

The bottom panel of Figure~\ref{fig:fractionPrior} shows the expected statistical uncertainty in the event rate predicted by the population simulation, for simulations with a total number of samples of $N = 10^4 $ and $ 10^6 $. By statistical uncertainty we mean the uncertainty that arises from using a finite number of samples. In standard Monte Carlo simulation this is also sometimes referred to as Poisson error, given for traditional sampling by $1 / \sqrt{N_{\rm{T}}}$. We estimate the minimum \AISs{} uncertainty in Fig.~\ref{fig:fractionPrior} as $1/\sqrt{N_{\rm{T}}}$, where ${N_{\rm{T}}}$ is now the total number of objects of interest found jointly in the exploration and refinement phases. The uncertainty as computed through Eq.~\ref{eq:variance} will be slightly greater, since the distribution of weights means that the effective sample size of the target population is lower than ${N_{\rm{T}}}$. These analytic estimates appear to be consistent with our numerical calculations (see also Section~\ref{subsec:VerificationCOMPASefficiency}). 

This is not the physical uncertainty since the model used for the simulation might still be wrong. 
In practice the efficiency in the refinement phase will not be perfect, i.e., not all samples drawn in that phase will find an outcome from the target population.  So the expected uncertainty from \AISs{} will lie in  the shaded region shown in Fig.~\ref{fig:fractionPrior}.  \AISs{} efficiency gains will be greatest for rare events and large $N$, allowing a greater fraction of time to be spent in the efficient refinement phase.

Comparisons to observational data will typically be made using distribution functions of predicted quantities (e.g., component masses), not just event rates. We later demonstrate the improvements provided by our algorithm for predictions of distribution functions.  Nonetheless this overall decrease in statistical rate uncertainty for fixed sample number in a simulation is indicative of the improvements enabled by applying \AISs \  to a target population.  

Figure \ref{fig:gain-for-imperfect-efficiency} shows the increase in the number of simulated binaries of interest versus traditional Monte Carlo sampling from a birth distribution for a simulation with fixed $N = 10^6$.  The efficiency in the refinement phase is not known in advance.   We show predictions for a refinement phase efficiency of 1 and 0.1; as long as the total number of successful samples is dominated by those drawn during the refinement phase, the maximum possible gain is roughly proportional to the refinement phase efficiency.  The value for the efficiency of the refinement phase varies between $\sim 3.4 \cdot 10^{-2} $ and 3.7 $\cdot 10^{-1}$ in our example astrophysical simulations (see Section \ref{sec:Results} and Table~\ref{tab:comparison2}). 


\section{Results}
\label{sec:Results}

In this section we demonstrate the power and advantages of \AISs.  Our algorithm could be applied to many sampling routines, but the illustration here uses the binary population synthesis code {\sc{COMPAS}}  \citep{stevenson2017formation, 2018MNRAS.477.4685B, 2018MNRAS.481.4009V}.     The physical assumptions and parameter settings we adopt are briefly summarised in Appendix~\ref{app:COMPASdetails}. 
 
We combine \AISs \ and \textsc{COMPAS} to model six different target populations. Four are simulations of subtypes of DCOs that merge in a Hubble time:  (1) all DCO mergers (i.e., BH--NS, NS--NS and BH--BH), (2) BH--BH mergers, (3) NS--NS mergers and (4) BH--NS mergers. Additionally we model  two simulations of extremely rare events by focusing on a subset of the above, namely (5) BH--BH mergers with total system masses in excess of $m_{\rm{tot}} \geq 50 M_{\odot}$   and (6) NS--NS mergers that merge within $t_{\rm{c}} \leq 50$ $\rm{Myrs}$ from the moment of the DCO formation (where $t_{\rm{c}} $ is the coalescence time). A summary of the results for these simulations can be found in  Table \ref{tab:comparison2}.  

In this section we present detailed results from simulations 1--4, as these target populations are those most commonly discussed in the literature. We just present the key findings for simulations 5 and 6.  For each target population we compare a simulation using our sampling algorithm to one which uses birth distribution Monte Carlo sampling, which for conciseness we will typically call traditional sampling.  Both the \AISs \ and the traditional simulations sample $N = 10^6$ initial binaries. 

The overall gain that is obtained when using \AISs{} depends on the simulation and the initial efficiency of the `traditional' method that the algorithm is compared with. For example, the choices for the initial parameter space can change how much the sampling can be improved when using \AISs. We use settings that are commonly used in population synthesis studies of DCO mergers. 

The remainder of the section is structured as follows. Section~\ref{subsec:VerificationCOMPASefficiency} demonstrates the increased efficiency of \AISs{} at finding binaries from the target population. Section~\ref{subsec:VerificationCOMPASspeedUp} discusses how that increased efficiency can be used to speed up simulations. Section~\ref{subsec:BHNSimprovement_moreHits} shows how our algorithm produces better resolution of the target population. Section~\ref{subsec:showcaseBHNS-smallerVariance} describes how our sampling method leads to smaller statistical uncertainties in predicted population distribution functions. Section~\ref{subsec:results-tails-of-distributions} shows how \AISs{} becomes even more important for recovering tails of distribution functions and when considering observational bias. 
Section~\ref{subsec:handling-bifurcations} discusses how \AISs{} handles well the bifurcations and discontinuities in the binary population synthesis parameter space. 

\begin{figure*}
	\includegraphics[width=1\textwidth]{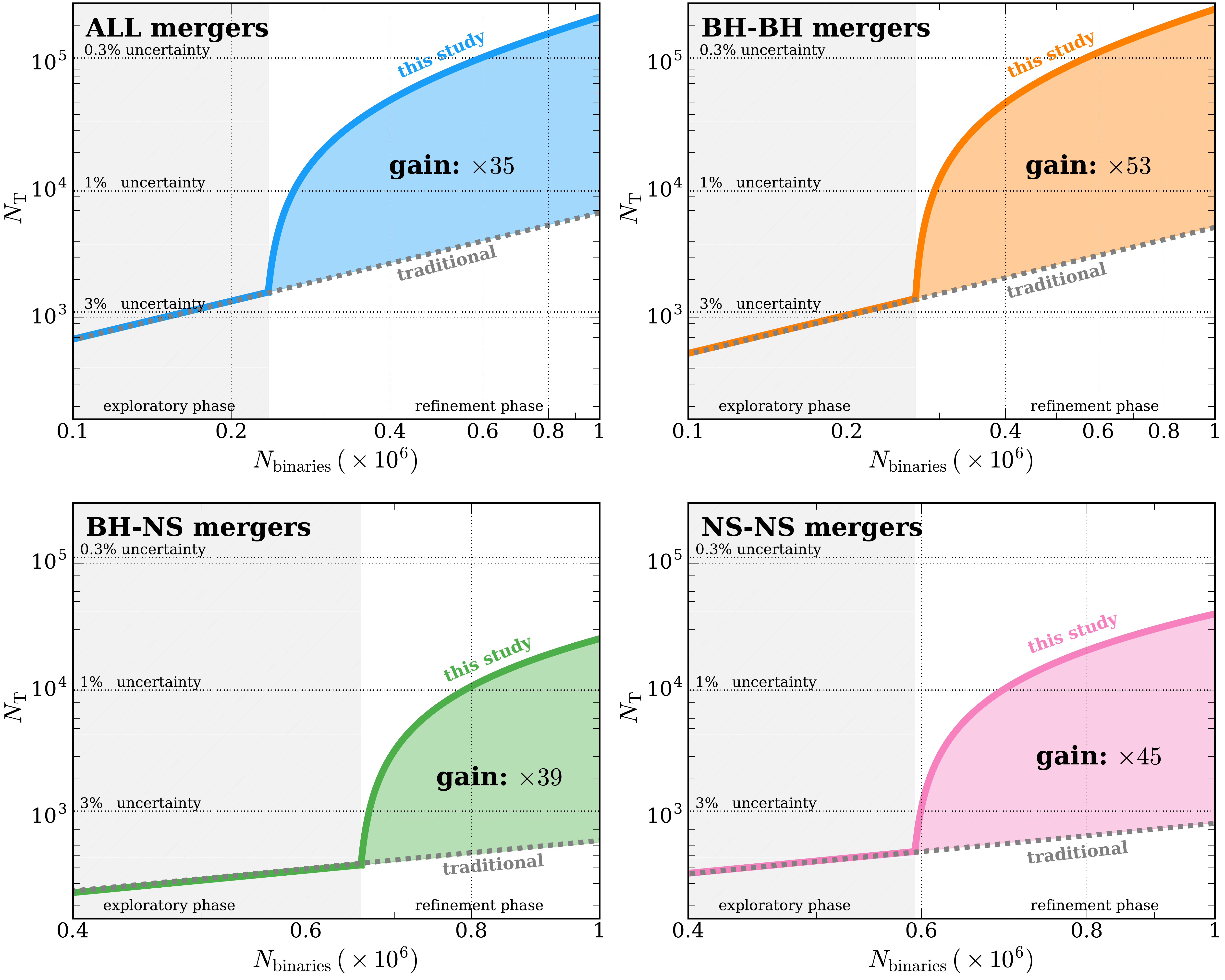} 
    \caption{The number of simulated binaries $N_{\mathrm{T}}$  falling into the target population  as a function of the total number of binaries $N_{\text{binaries}}$ sampled for the traditional sampling method (gray dashed line) and the sampling method presented in this study (solid coloured line). The four panels show the simulations for each of the four target sub-populations. In each panel the duration of the exploratory phase is shown with a hashed gray area.  In the background the standard Poisson fractional uncertainties  of $0.3, 1$ and $3\%$ are shown with  dashed lines. } 
    \label{fig:NbinariesVsNHits}
\end{figure*}

\subsection{On the gain of generating binaries of the target distribution}
\label{subsec:VerificationCOMPASefficiency}
We find that the number of binaries of these target populations increases by factors of about 25 -- 200 when using our \AISs \ sampling algorithm compared to simulations with traditional sampling. 
The panels in Fig.~\ref{fig:NbinariesVsNHits}  showcase this by presenting the number of systems formed from the target population as a function of total number of sampled binaries, for both our sampling method and traditional birth distribution Monte Carlo sampling. For these four target populations the gains are between $\sim$35--55.  For the two additional extremely rare target populations we find gains of $24$ and $203$. The gains are also shown  in the last column of Table~\ref{tab:comparison2}, and Figure \ref{fig:gain-for-imperfect-efficiency}.

\begin{table*} 
\centering
\begin{tabular}{|c|l|c|c|c|c|c|c|c|}
\hline
nr & Target subpopulation     		    &   $f_{\rm{expl}}$  &   {efficiency}    		&  { efficiency}  & gain & $N_{\text{T}}$ & $ N_{\text{T}}$ &   gain  	\\
&	       &			                        & 	exploratory  		          &	  refinement     & refinement         & traditional  & \AISs{} & 	 overall\\ \hline
1 & All DCO mergers in a Hubble time   &  \fexplOne  & \EffExplOne  & \EffRefOne  & 45$\times$ & \NxMCOne   & \NxAISOne & \gainOne  \\
2 & BH--BH mergers in a Hubble time    &  \fexplTwo  & \EffExplTwo  & \EffRefTwo & 70$\times$ & \NxMCTwo   & \NxAISTwo & \gainTwo  \\ 
3 & BH--NS mergers in a Hubble time     &  \fexplThree  & \EffExplThree & \EffRefThree & 116$\times$ & \NxMCThree   & \NxAISThree & \gainThree  \\ 
4 & NS--NS mergers in a Hubble time     &  \fexplFour  & \EffExplFour  & \EffRefFour & 108$\times$ & \NxMCFour   & \NxAISFour & \gainFour  \\ \hline
5 & BH--BH mergers $m_{\rm{tot}} \geq 50 \, \rm{M}_{\odot}$      &  \fexplFive  & \EffExplFive &  \EffRefFive & 651$\times$ & \NxMCFive   & \NxAISFive & \gainFive  \\
6 & NS--NS mergers with $t_{\rm{c}} \leq 50 \, \rm{Myr}$   &  \fexplSix  & \EffExplSix  & \EffRefSix & 99$\times$ & \NxMCSix   & \NxAISSix & \gainSix  \\ \hline
\end{tabular}
\caption{Summary of the results from six  target populations that are modelled in this paper to demonstrate our \AISs{} algorithm.  We list the fraction of samples spent in the exploratory phase, $f_{\rm{expl}}$, and the efficiency of finding `hits' in the exploratory and refinement phases. The gain in refinement is the ratio between the efficiency of finding samples of the target population during the refinement phase  of  \AISs{} and traditional sampling (where the efficiency of traditional sampling is equal to the efficiency of the \AISs{} exploratory phase).   $N_{\text{T,traditional}}$ and  $ N_{\text{T,\AISs{}}}$ represent the total number of systems of interest that are found by the end of the simulation (using a total of $10^6$ samples). The last column is the overall gain that we found when using \AISs{} compared to traditional Monte Carlo sampling from the birth distributions, which is defined by the ratio   $ N_{\text{T,\AISs{}}}$ / $N_{\text{T,traditional}}$.  \href{https://doi.org/10.5281/zenodo.3387651}{\color{linkcolor}\faBook}}
\label{tab:comparison2}
\end{table*}

At the beginning of each simulation, during the \AISs \ exploratory phase, the two sampling methods produce similar number of binaries of interest (i.e., hits), only different by random chance.  The duration of that exploratory phase is determined by $f_{\rm{expl}}$, as derived in Section \ref{subsec:durationExploratory}. For our simulated target populations, using $N=10^6$ samples, $f_{\rm{expl}}$ ranges between $\approx 0.2$ and $0.8$ (see $f_{\rm{expl}}$ in Table~\ref{tab:comparison2}).   The algorithm then switches to the more focused refinement phase, using the information from the hits found during the exploratory phase.   During this refinement phase our sampling algorithm is 45 -- 650 times more efficient at finding hits (see the sixth column in Table~\ref{tab:comparison2}).    

The difference in efficiency gains between the populations originates mostly from two effects. First, the different rarenesses of the target populations (e.g., for these assumptions BH--NS mergers are more rarely produced than BH--BH mergers) influences how much the efficiency increases during the refinement phase and also the duration of the exploratory phase. Both are important factors for determining the overall gain in efficiency.   Second, the structure of the output surfaces influences how well the Gaussian mixture distribution covers the regions of interest in the output space. A more stochastic or discontinuous output space (e.g., many small islands of hits) will lead to smaller  efficiencies in the refinement phase of \AISs. This effect is most noticeable in the different gains between the BH--BH systems with total masses over 50 $\rm{M}_{\odot}$ and the NS--NS systems that merge within 50 $\rm{Myrs}$. Supernova remnants receive a random natal kick in our simulations.   These kicks induce stochastic and discontinuous behaviour into the output surfaces, leading to lower refinement-phase efficiency and gain.   The kicks are typically larger for NSs than BHs, and so affect NS--NS simulations the most.   Conversely, the gain is greatest for the BH--BH merger populations which are least affected by these stochastic kicks (as shown in Figs.~\ref{fig:fractionPrior} and \ref{fig:gain-for-imperfect-efficiency}).

The largest overall gain shown in Table~\ref{tab:comparison2}, and Fig. \ref{fig:gain-for-imperfect-efficiency} is for modelling a BH--BH population with total mass over 50 $\rm{M}_{\odot}$. For this population it would be possible to increase the efficiency of Monte Carlo sampling from the birth distributions by making thoughtful changes to the boundaries of the initial parameter space, so the factor of $\approx$200 improvement we show for \AISs{} is higher than would arise when comparing to more carefully-targeted use of standard Monte Carlo sampling in this case.  However, well-informed choices in the initial parameter space would also benefit \AISs{} by increasing the efficiency of the initial exploration phase. Moreover, one-dimensional cuts to the parameter space would become increasingly inefficient when sampling in higher dimensions.  \AISs{} automatically finds the regions of interest, and avoids the risk of incorrect choices in restricting the initial parameter space.

\begin{figure}
	\includegraphics[width=1\columnwidth, trim={0cm .5cm 0 0},clip]{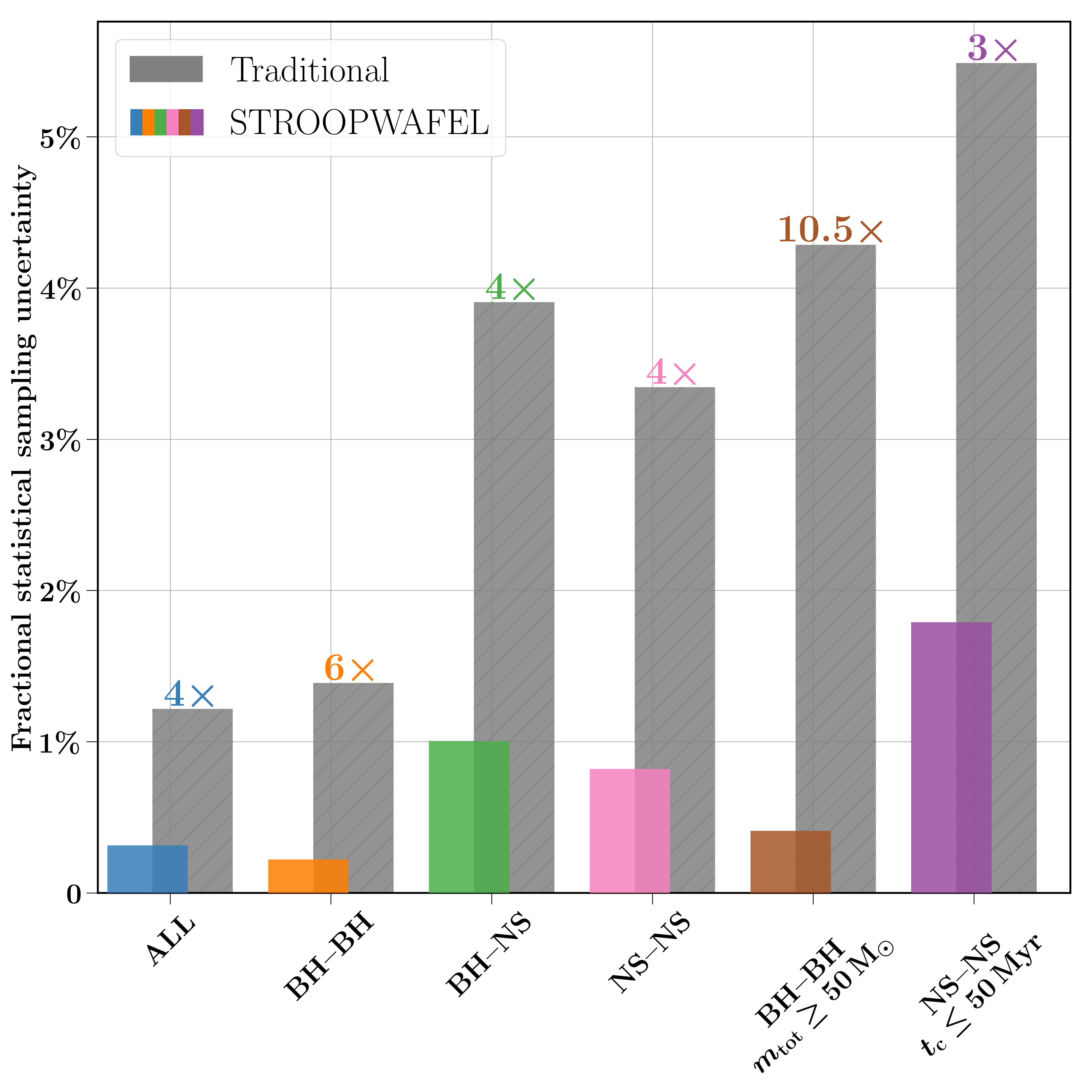}
    \caption{Sampling uncertainty estimate from each simulation of the target population. Gray bars show the uncertainty from traditional sampling methods whereas the coloured bars show the uncertainty from our sampling method \AISs. All simulations use a total of $10^6$ samples. The number shown on top of the traditional bar shows the factor in decrease in uncertainty from \AISs{} compared to traditional sampling for that simulation. \href{https://doi.org/10.5281/zenodo.3387651}{\color{linkcolor}\faBook}} 
    \label{fig:RatesUncertainties}
\end{figure}

The increase in the number of events  decreases the sampling uncertainty in the predicted event rates.  Although the standard uncertainty from Poisson noise decreases with the square root of the number of target systems found, i.e., as $1 / \sqrt{N_{\rm{T}}}$, in our weighted sampling case it also depends on the variance in the weights (see Eq. \ref{eq:variance}).
We find that our sampling algorithm  results in $\approx 3$ -- $10.5$ times smaller sampling uncertainties compared to traditional sampling for the same total of samples $N = 10^6$. This is presented in Figure~\ref{fig:RatesUncertainties}, which shows the fractional statistical uncertainty estimate on the rate estimate, i.e.,$\sqrt{{\mathbb{V}[\rate_{\text{T}}]}}  / {\mathbb{E}[\rate_{\text{T}}]} $  from each simulation. 

\subsection{Speeding up simulations}
\label{subsec:VerificationCOMPASspeedUp}

Instead of using \AISs \ to obtain more information from a simulation with the same number of samples, one could alternatively aim for a certain precision in the predicted event rates. In that case \AISs \ can be used to speed up the simulation, since this precision will be reached using a fraction of the number of samples required when using traditional sampling.   Traditional sampling would require 25 -- 200 more simulations than \AISs \ to achieve the same number of target binaries, and a factor of around 10 -- 100 times more simulations to achieve the same rate estimate precision (these speed-up factors differ because the statistical uncertainty depends on the variance in the weights as well as the number of target samples).

The speed-up factor further depends on the computational cost of simulating samples from the chosen distribution.  It might be that the binaries of interest require more or less computational time than other binaries. Therefore the speed-up when using the adaptive distribution $Q$ depends on the science case of interest. In the simulations performed for this study the average computational cost (in CPU time) of simulating typical individual binaries sampled from the adaptive distribution $Q$ was up to a factor of 2 smaller than for individual binaries sampled from the birth distribution.  Therefore, the total speed-up was up to another factor of 2 larger in our simulations than from more efficient sampling alone.

More generally, we note that the gain or relative speed-up from using \AISs{}  will depend on the target population and the traditional method with which it is compared.  First of all, the speed up from \AISs{} will generally be greater (smaller) if the target population is more (less) rare. This is shown in Fig.~\ref{fig:gain-for-imperfect-efficiency}. Equivalently, if one chooses a larger initial parameter space (e.g. sampling $m_{\rm{1,i}}$ from $[1,150] \,\rm{M}_{\odot}$ instead of  $m_{\rm{1,i}}$ from $[5,150] \,\rm{M}_{\odot}$ used here), the gain would have been larger as the event of interest becomes rarer (assuming no binaries in the extended range form a binary of the target population).  Secondly, in some binary population synthesis studies the primary mass is sampled uniformly in  $\log \, m_{\rm{1,i}}$ space. This is a form of importance sampling. The gain of using \AISs{} (with uniform sampling in $\log m_{\rm{1,i}}$ during the exploratory phase) could be lower than gains without this importance sampling if importance sampling makes the traditional Monte Carlo more efficient. Nevertheless, we would still expect a significant gain from  \AISs{} - especially if using that form of importance sampling in the exploratory phase significantly decreases the duration of this phase. 

\subsection{Mapping the parameter space with higher resolution}
\label{subsec:BHNSimprovement_moreHits}

\begin{figure*}
	\includegraphics[width=.95\textwidth]{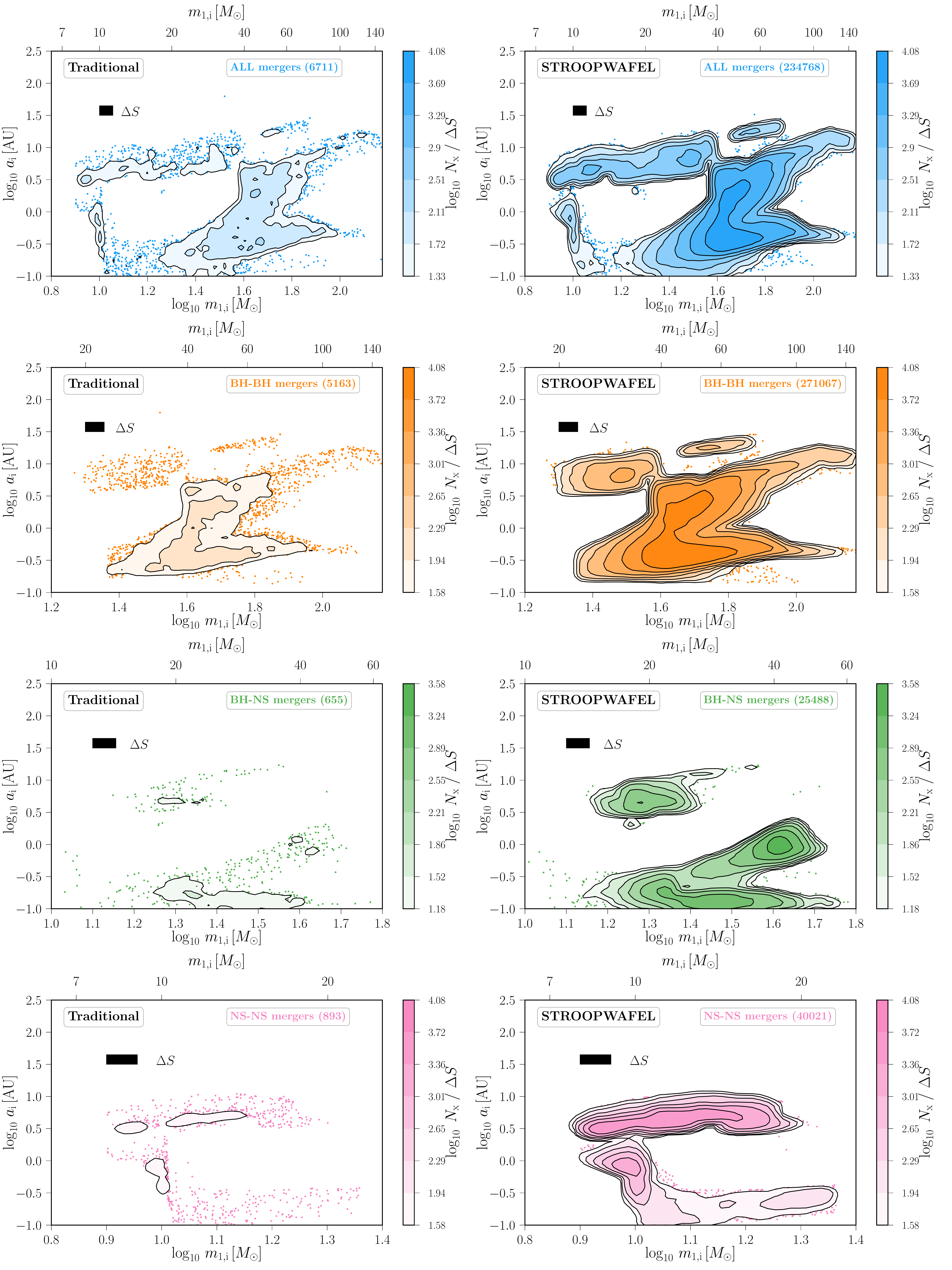}
    \caption{Contour plots of the locations in $\log \ m_{\rm{1,i}}$ and $\log \, a_{\rm{i}}$ space of the hits $\boldsymbol{{x_{\rm{T}}}}$ (i.e., binaries of the target population) found in each simulation when using traditional birth distribution Monte Carlo sampling (left panels) and the sampling method \AISs \ developed in this study (right panels). Contours represent a constant density of binaries of the target population found per unit area in $\log \ m_{\rm{1,i}}$ -- $\log \, a_{\rm{i}}$ space. The colour gradient indicates the number of samples per area $\Delta S$, the size of which is shown with a black rectangle.  If the density is below the level of our lowest contour we plot the individual points. The four different panels from top to bottom represent the first four target populations shown in Table \ref{tab:comparison2}.  The total number of hits $N_{\text{T}}$ found in each simulation is quoted in  parentheses. The metallicity assumed in all simulations is $Z = 0.001$.    }
    \label{fig:ContourPlotInitial}
\end{figure*}

\begin{figure*}
	\includegraphics[width=0.95\textwidth]{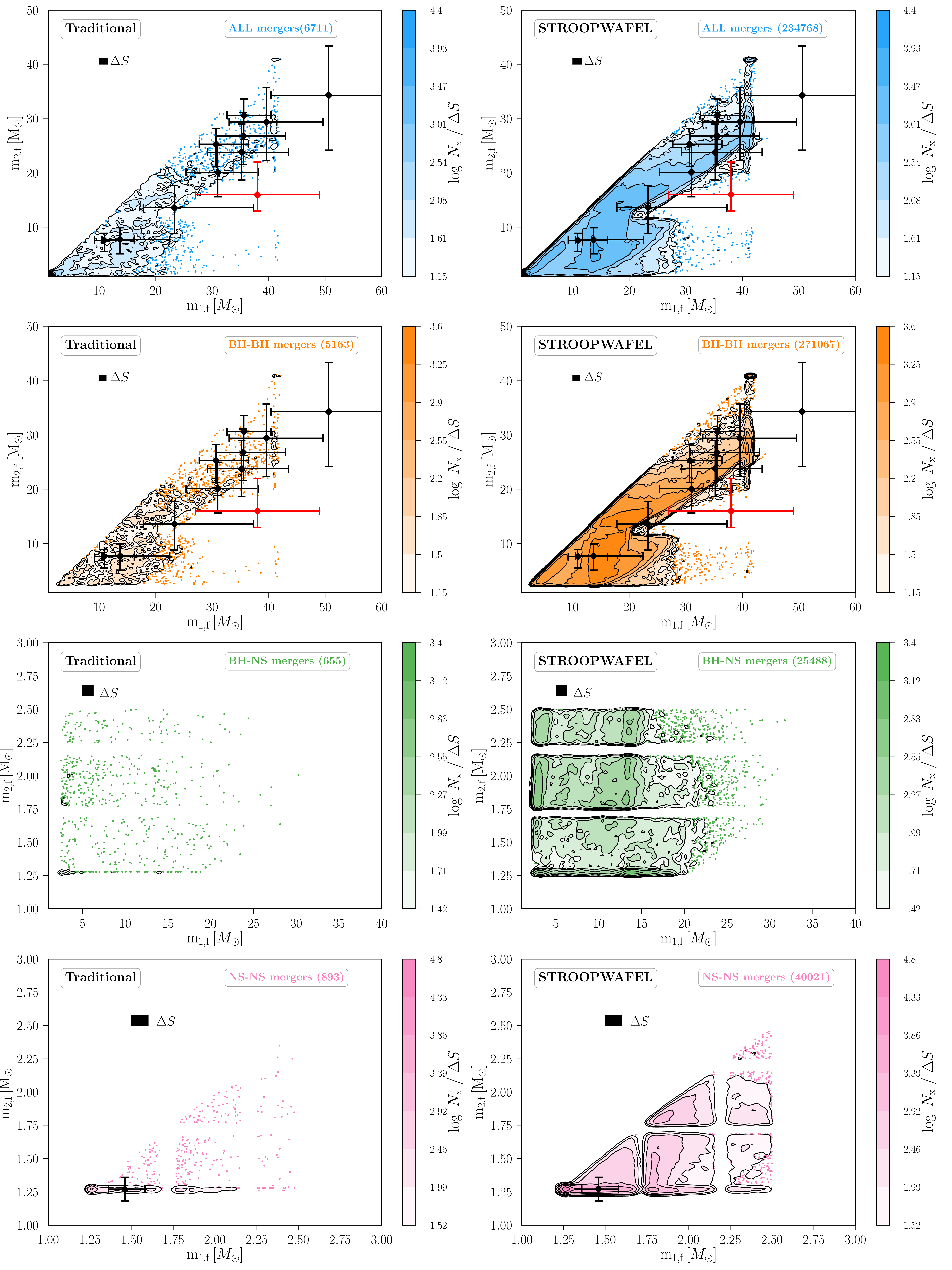}
    \caption{Similar to Figure~\ref{fig:ContourPlotInitial} but now for the output parameters: the final compact object masses $\rm{m_{\rm{1,f}}}$ and $\rm{m_{\rm{2,f}}}$ of the DCO.  We overplot the gravitational-wave observations from  O1 and O2  from \citet{2018arXiv181112907T} in black and  from \citet{2019arXiv190210331Z} in red. Error bars indicate 90$\%$ credible regions around the median.  The metallicity assumed in all simulations is $Z = 0.001$; selection effects of gravitational-wave detectors are not accounted for.}
    \label{fig:ContourPlotFinal}
\end{figure*}

The increase in computational efficiency from \AISs{} leads to  finding substantially more events of the target population which  naturally enables a much higher resolution mapping of both the input and output parameter spaces.   Figures~\ref{fig:ContourPlotInitial} and~\ref{fig:ContourPlotFinal}   show  examples of how the parameter space is explored in far greater detail using our sampling method compared to traditional birth distribution Monte Carlo sampling.  

Figure~\ref{fig:ContourPlotInitial} shows the location of the target population in the initial parameter space of primary mass  $m_{\mathrm{1,i}}$ and separation $\log  \, a_{\rm{i}}$ at birth.  With our sampling method we obtain more detailed contours and more contour levels that map the initial parameter space with higher resolution. This leads to better knowledge of the initial conditions of a binary system that yield a binary of the target population. Physically the structures seen in the input parameter space correspond to the assumed physics of the different formation channels leading to compact-object mergers. More details are discussed in \citet[]{stevenson2017formation,2018MNRAS.481.4009V}.  

Meanwhile, Fig.~\ref{fig:ContourPlotFinal} shows the higher resolution mapping from \AISs \ for the output space of the final  masses of the compact objects  $\rm{m_{\rm{1,f}}}$ and $\rm{m_{\rm{2,f}}}$ in each DCO. We plot on top the gravitational-wave events found from O1 and O2 data from  \citet{2018arXiv181112907T} and \citet{2019arXiv190210331Z}\footnote{Publicly available data can be found at \url{https://www.gw-openscience.org/catalog/GWTC-1-confident/html/}.}. The simulations with our \AISs \ algorithm again yield higher resolutions and more systems of the target populations in the regions overlapping with the observations. This is important in order to compare  observations and theory and test the physical assumptions in our models. 

Figure~\ref{fig:ContourPlotFinal} shows that even in the \AISs{} simulation, there are relatively few samples consistent with the 90$\%$ credible regions for some of the highest mass gravitational-wave events  observed in O1 and O2.  The samples shown here are not weighted for the sensitivity of gravitational-wave interferometers. In addition, they correspond to a particular model chosen in our simulation, and in practice many variations of this model need to be explored  for a meaningful comparison with observations \citep[e.g.,][]{2018MNRAS.477.4685B}. For example, the metallicity in this model is fixed to $Z = 0.001$ and we expect to form more massive black holes at lower metallicities \citep[e.g.,][]{2015MNRAS.451.4086S,2016Natur.534..512B, 2019arXiv190608136N}. In addition, the most massive BH--BH mergers may have formed through a different formation channel than the classical isolated binary evolution via the common-envelope phase which is simulated with \textsc{COMPAS}. One example of such a different formation channel is chemically homogeneous evolution, explored by  \citet[][]{10.1093/mnras/stw379, 2016MNRAS.460.3545D, 2016A&A...588A..50M}. Another example  is dynamical formation of BH--BH mergers in a dense stellar environment (see \citealt{2018arXiv180605820M,2018arXiv180909130M} for reviews). Another possible explanation is that some of these highest-mass events are instead from instrumental origin as they also have a relatively high false-alarm-rate \citep{2018arXiv181112907T, 2019arXiv190210331Z}.

The BH--NS and NS--NS panels in Fig.~\ref{fig:ContourPlotFinal}  show discontinuities in the NS remnant mass, with obvious gaps in the NS mass range. 
This is a consequence of discontinuities in the \texttt{delayed} \citet{2012ApJ...749...91F} model describing the mapping from the carbon--oxygen core mass to the remnant (NS) mass, which is used in the Fiducial COMPAS model. Although such discontinuities may be physical, this mapping does not reproduce the mass distribution of Galactic NS--NS binaries \citep{2018MNRAS.481.4009V}.  For the purposes of this paper, our main message is how such features in model populations can be clarified by improved sampling. 

\subsection{Smaller variances in distribution functions}
\label{subsec:showcaseBHNS-smallerVariance}

The most important consequence of the increase in sampling efficiency enabled by \AISs \ is  that it leads to a significant decrease in the statistical uncertainty of the predictions for the output parameter spaces. That is, it improves the precision in the predicted population observables. 

\begin{figure*}
	\includegraphics[width=1\textwidth]{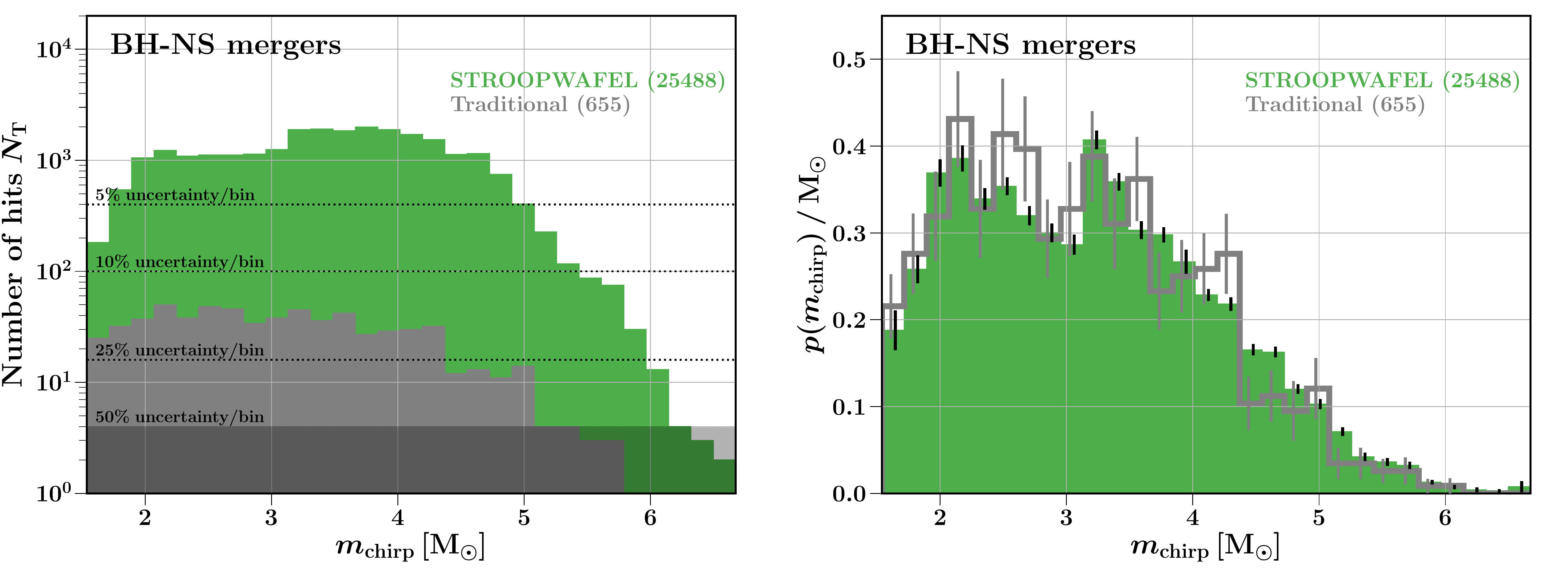}
    \caption{\textbf{Left panel:} histograms of the number of target BH--NS binaries $N_{\rm{T}}$ found per chirp mass bin $m_{\rm{chirp}}$ for traditional Monte Carlo sampling (grey) and the \AISs{} sampling method presented in this study (green). The standard Monte Carlo fractional uncertainties (i.e. Poisson noise) are shown with dashed lines in the background. We mark everything below 4 events (i.e. $50\%$ uncertainty) as statistically insignificant as it is consistent with no hits within 2 standard deviations. \textbf{Right panel:} the BH-NS chirp mass probability distribution; {\AISs{} results have been re-sampled with weights from Eq.~(\ref{eq:weightsMixture}).}  The metallicity assumed in the simulations is $Z = 0.001$. The bin width is approximately $ 0.2 M_{\odot}$  and is constant between the traditional and \AISs{} algorithm. \href{https://doi.org/10.5281/zenodo.3387651}{\color{linkcolor}\faBook}}
    \label{fig:mtotHistograms}
\end{figure*}

Figure~\ref{fig:mtotHistograms} illustrates this improvement. The left panel in Fig.~\ref{fig:mtotHistograms}  shows the number of binaries of the target population  $N_{\rm{T}}$ found within a certain chirp mass  bin  for the BH--NS simulation.  The chirp mass $m_{\rm{chirp}} = (m_{\rm{1,f}} m_{\rm{2,f}})^{3/5}  /  (m_{\rm{1,f}} + m_{\rm{2,f}})^{1/5}  $  is a combination of the masses of the DCOs that is particularly accurately measured with gravitational-wave observations.  For the same histogram bin widths, i.e., the same DCO chirp mass resolution, our improved sampling leads to more binaries of the target distribution per bin, and hence yields smaller fractional sampling uncertainties for each histogram bin.  This is shown in the right panel of Fig.~\ref{fig:mtotHistograms}, which displays the normalised chirp mass distributions from traditional and \AISs{} sampling. The error bars showing the statistical sampling uncertainty on each bin are much smaller for our sampling algorithm, leading to better predictions for the distribution functions. 

\subsection{Recovering tails of distribution functions}
\label{subsec:results-tails-of-distributions}
\begin{figure*}
    \includegraphics[width=1\textwidth]{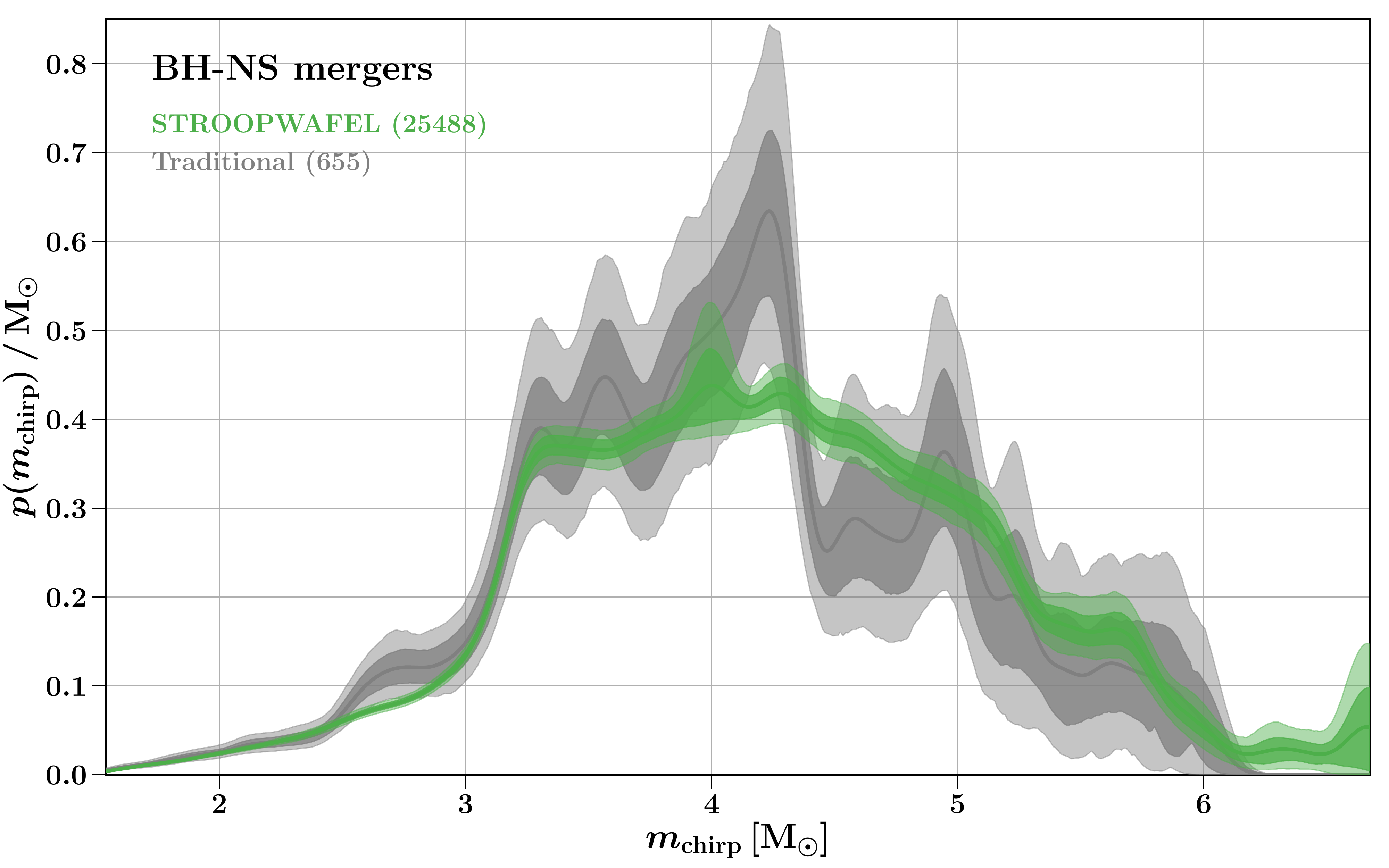}
    \caption{Predicted distribution of the chirp mass of the merging BH--NS population using \AISs{} (green) and traditional (grey) sampling. In both cases the simulation uses $N= 10^6$ samples and the distributions are weighted by the sensitivity of gravitational-wave interferometers using   \citet{2017ApJ...851L..25F}. Shaded regions show the  $1$- and $2$--$\sigma$ confidence intervals which are calculated by bootstrapping the samples 1000 times.   
This distribution is for a particular set of model assumptions, including a single metallicity $Z = 0.001$, and an integration over a metallicity-dependent cosmic star formation history is required for comparisons with observations. The same \texttt{scipy} kernel density estimator smoothing with a dimensionless kernel density estimator factor of about $0.1 $ is used for traditional and \AISs{} distributions (see also Appendix \ref{app:KDEbandwidth}).}
    \label{fig:Mchirpkde}
\end{figure*}

The need for more efficient sampling algorithms in binary population synthesis simulations of gravitational-wave source progenitors becomes even more evident when we consider the observational biases of gravitational-wave detectors.   These biases, which generally favour high-mass systems with greater gravitational-wave amplitudes, over-emphasise the rare and frequently under-sampled tails of the simulated distributions.  An example is shown in  Fig.~\ref{fig:Mchirpkde}, where we plot the predicted distribution of  chirp masses for  BH--NS systems estimated  using traditional birth distribution Monte Carlo or by using the  \AISs{} algorithm. The shown distributions are  weighted by the sensitivity of gravitational-wave interferometers, approximated as a bias dependent on the primary DCO mass $\propto  m_{\rm{1,f}}^{2.2} $ \citep{2017ApJ...851L..25F}. We also show  $1$- and $2$--$\sigma$ confidence intervals which are calculated by bootstrapping the samples 1000 times. Our algorithm produces much smoother distribution predictions with much smaller sampling uncertainties compared to traditional sampling methods for the same number of samples simulated. 
In particular, Figure~\ref{fig:Mchirpkde} demonstrates that simulations using the traditional Monte Carlo sampling from the birth distributions under-sample the high-mass end of the population.  This will be particularly significant when comparing population models to observations.


Figure~\ref{fig:Mchirpkde} corresponds to a particular model choice; variations of the model have to be considered in order to compare with observations. The displayed distribution is from a simulation at a single metallicity of $Z= 0.001$, while a range of metallicities will contribute to the observed BH--NS merger population. An integration over the metallicity-dependent cosmic star formation history is therefore required.  The properties of BH--NS mergers will be explored with \textsc{COMPAS} in future work. 

\subsection{Handling bifurcations and stochasticity}
\label{subsec:handling-bifurcations}
One of the most important results is that our sampling algorithm \AISs{} handles well the bifurcations and stochasticity that naturally occur in the parameter spaces of binary population synthesis simulations. This discontinuous behaviour is visible in Figs.~\ref{fig:ContourPlotInitial} and~\ref{fig:ContourPlotFinal} by the disconnected contours and the offset of the location of some individual points from those contours. Such offset points physically relate to extremely rare formation channels or tails of distribution functions, while the ridges in the birth parameter space relate to bifurcations in the fate of the binary.  Not only does \AISs{} recover the irregularly shaped structures in the parameter space, our algorithm also finds these more scattered points. 

\section{Discussion}
\label{sec:discussion}

We have demonstrated that the performance of \AISs \ is substantially superior to traditional Monte Carlo sampling from the birth probability distributions.    For the types of rare events simulated in Section \ref{sec:Results}, the gain is already so large that the current implementation of our algorithm can contribute to drastic speed-ups of binary population synthesis simulations.  Hence we have postponed some natural improvements to \AISs \ until later, but we discuss them here.  After those, we discuss additional potential applications for our algorithm.

\subsection{The exploratory phase}
During the exploratory phase in \AISs \ the initial parameter space is sampled by drawing random binaries from the priors (as in traditional birth distribution Monte Carlo sampling) until $N_{\text{expl}}$ events of interest are found. There are several features of the exploratory phase that could be optimised and improved.

\begin{itemize}
\item  We now use sampling from the birth distribution $\pi$ for drawing the random binaries in the exploratory phase. Future improvements of \AISs \ could use more efficient sampling algorithms in the exploratory phase. Examples include (1) using importance sampling in the exploratory phase when there is an existing guess at a more efficient sampling distribution, or (2) implementing techniques such as Latin hypercube sampling  \citep[LHS,][]{10.2307/1268522, eglajs1977new, iman1980latin, iman1981approach}. LHS is a  Monte Carlo method that  generates near-random samples which are more equally distributed throughout the initial parameter space by placing only one sample in each row and column of the Latin square\footnote{A Latin square of order $n$ is an arrangement of $n$ different variables in a $n \times n$ array such that each variable occurs exactly once in every row or column \citep{euler1782recherches}.}. By doing so, it could improve the sampling in the exploratory phase as the probability of the randomly drawn samples being clustered decreases slightly.
\item The duration of the exploratory phase is now determined with $f_{\rm{expl}}$, which is optimised for the uncertainties on the rates of the target distribution. A future improvement would be to determine $f_{\rm{expl}}$ based on the uncertainty in the simulated output distribution function. See also Sect.~\ref{subsec:discussion-adaptingUncertainty}. 
\item The exploratory phase duration is optimised under the simplifying assumption that the instrumental distribution will match the target distribution except for some missing regions in parameter space.  The optimisation could also consider the level of fluctuation in the instrumental sampling distribution (i.e., the variance in the weights).
\item If the structures in the parameter space have a known minimum volume, we could use this to derive a better informed estimate for the uncertainty contributing from the probability of missing  such structures in the exploratory phase.  This seems unlikely to apply to binary-star population synthesis, but might be relevant for other applications of \AISs.
\end{itemize}

\subsection{The refined sampling phase}
\label{subsec:discussion-refinedSampling}

The Gaussians which are used to form the instrumental distribution are currently constructed using diagonal covariances ($\Sigma_k$).    These then remain unchanged throughout the refinement phase of adaptive importance sampling -- even though much more information becomes available about the distribution of hits in the initial parameter space.   A potential future improvement is to update the instrumental sampling distribution during the refinement phase.    In principle this might be done locally, with only the samples drawn from each of the individual Gaussians used to update the corresponding element of the instrumental distribution.  Doing so would avoid a potentially expensive nearest-neighbour search, as the tree is automatically built for free by the sampling already being performed. See also for example the AMIS algorithm described in  \citet{cornuet2012adaptive}.  Adaptive distribution choices beyond a mixture of Gaussians could also be explored.

\subsection{Adapting to uncertainty in distribution functions}
\label{subsec:discussion-adaptingUncertainty}

Observational selection effects must be applied to model predictions in order to statistically compare models against observations.  These selection effects are generally applied after population synthesis models are generated, and may place significant weight on rarely-formed systems in the tails of output distribution functions (e.g., higher-mass DCO binaries).  Even though  \AISs \ can greatly improve the overall number of systems produced from a simulation, there may still be relatively few systems in these tails.

In principle, we could tune \AISs \ to produce a model population weighted towards any observational population distribution, i.e., optimising for observational selection effects and spending less time on systems which do not contribute to the observed sample. This can be achieved by incorporating selection effects directly in the instrumental distribution rather than applying them to \AISs \ outputs.  This can be practically implemented by changing the instrumental distribution weights. At the moment all Gaussians contribute equally to the mixture distribution with a weight $ 1 / N_{\rm{T, expl}}$ but instead the contribution of each Gaussian can be weighted with the probability of observing the system to focus the simulation on systems that are more likely to contribute to the observable population.  An extreme example of this approach would be re-defining the target population to be an even rarer subset of the initial target population, e.g., tails of a distribution function. \AISs \ could also be used to sample from regions of the initial parameter space giving rise to properties consistent with specific observed systems \citep[see also][]{andrews2017dart_board}.

The current implementation of \AISs \ might be thought of as something like adaptive mesh refinement, familiar from hydrodynamics, applied to the phase space of binary population synthesis.   This potential future development of \AISs \ would be refining in the space of predicted observables.   This development of  \AISs \ could naturally be applied to modelling any population for comparison to observations, not only intrinsically rare populations.

\subsection{\AISs{} in higher dimensions}
\label{subsec:DiscussionHigherDim}

The demonstrations in this paper have all used \AISs \ to sample in a three-dimensional birth parameter space of the two component masses and the initial orbital separation.  \AISs \ can be readily applied to sample in more dimensions. The scaling parameter $\kappa$ may well have a different optimal value for higher dimensions, and should be investigated before being applied to higher dimensions, although we anticipate moving away from using diagonal covariances (see subsection \ref{subsec:discussion-refinedSampling}). Additional potential dimensions to add to the space of initial conditions include, e.g., initial compositions, the initial eccentricity of the system, or the spins of the stars.   Moreover, in \textsc{COMPAS} systems are labelled from the start with vectors representing normalised versions of the supernova kicks that will be applied during compact-object formation \citep[see also, e.g.,][]{andrews2017dart_board}; each kick adds three dimensions to the parameter space. Potentially, applying \AISs{} to the kick vectors can be promising since the kick magnitudes and directions can significantly affect the fates of the systems. In our simulations this would be especially important to increase the gain in simulating NS--NS mergers, as the kicks contribute most to the current stochastic output surfaces for this target population.

\subsection{Combining  \AISs \ with MCMC or Gaussian process regression emulators on continuous spaces}
\label{subsec:DiscussionContinuousSpaces}
\AISs \ could be applied directly to the combined parameter space of initial parameters of an individual system (e.g., the initial masses and separations) and hyper-parameters describing the model assumptions (e.g., wind-driven mass loss rates, common-envelope physics).  

Alternatively, \AISs \ could be combined with other methods for exploring the parameter space, such as emulators based on Gaussian process regression \citep[see, e.g.,][]{2017IAUS..325...46B,2018arXiv180608365T}.  Some of the parameters map continuously to the output space, while others exhibit discontinuities.  \AISs \ distinguishes itself in handling bifurcations and stochastic output surfaces. On the other hand, emulators can be more efficient in sampling parameters that are smooth.  
Thus an intended future development is to combine \AISs \ with such methods to obtain the best overall efficiency. 

\AISs{} output samples could also be converted into probability distributions using Gaussian mixture models based on Dirichlet processes \citep{2018MNRAS.479..601D}.  

\section{Summary and conclusions}
\label{sec:conclusion}
We have presented a new sampling algorithm that aims to improve the efficiency of simulating rare events in astrophysical populations, and demonstrated its utility for binary population synthesis of gravitational-wave merger populations.  Our algorithm \AISs \  adaptively improves the sampling distribution to focus more computational time on the target population. Some key findings of our investigation are:

\begin{enumerate}
	\item Using \AISs \ we find a factor of about 25--200$\times$ more systems of interest in simulations of a certain length, as compared to Monte Carlo sampling from the birth distributions. To simulate the same number of events of interest with such commonly-used Monte Carlo sampling would require up to two orders of magnitude more computational time.  This gain  will improve binary population synthesis simulations by making it computationally feasible both to include more details of the relevant massive-star physics and to explore a greater number of variations of the physical assumptions of the model.\\
	
	\item The increase in efficiency of \AISs \ leads to higher-resolution mapping of both the input and output parameter space. This reduces the sampling uncertainty by factors of $\approx 3$ to 10 for our simulations with $10^6$ total samples. \\
	
	\item \AISs \ improvements are particularly significant when simulating extremely rare events or tails of distribution functions, such as the most massive BH-BH mergers or early NS-NS mergers.  \\ 
	
\item One of the core strengths of  \AISs{} is that it can handle well the bifurcations and discontinuities that naturally occur in the parameter spaces of binary population synthesis simulations. Such stochasticity often poses a challenge for applying sampling and emulation methods such as Markov Chain Monte Carlo and Gaussian process regression emulators that rely on smoothness to converge and produce independent samples. \\

\end{enumerate}

Future improvements to the \AISs \ algorithm  (discussed in Section \ref{sec:discussion}) should be able to further improve its performance. This could make it more realistic for next-generation binary population synthesis simulations to include detailed stellar evolution models whilst also exploring more variations in the model physics and assumptions.    Such improvements will help in comparing population models to population data, and so help to constrain the physics of evolutionary processes occurring on timescales too long to directly observe.

\section*{Acknowledgements}
We thank especially  T. Fragos, D. Sz\'{e}csi  and  M. Renzo for constructive discussions and comments. We also thank C. Berry, R. Willcox and M. Wassink for their helpful suggestions on this paper.   We thank D. Stops for technical support. We thank the anonymous referee for their helpful comments and careful feedback on this manuscript.
FSB, AVG, IM, SJ and JG thank the  Kavli Foundation, Niels Bohr institute and DARK Cosmology Centre in Copenhagen for their hospitality and for organizing the Kavli summer school in gravitational-wave astrophysics 2017 where part of this work has been performed.  
The work has been performed under the Project HPC-EUROPA3 (INFRAIA-2016-1-730897), with the support of the EC Research Innovation Action under the H2020 Programme; in particular, FSB gratefully acknowledges the support of the School of Mathematics, University of Edinburgh and Birmingham Institute for Gravitational Wave Astronomy,  University of Birmingham and the computer resources and technical support provided by EPCC. 
FSB also acknowledges support  through the McKinsey excellence grant and Kapteyn grant. 
SJ and IM thank the Aspen Center for Physics for hospitality whilst part of this work was performed, supported by National Science Foundation grant PHY-1607611. SJ is further grateful for partial support of this work via a grant from the Simons Foundation.
SdM and FSB acknowledge funding by the European Union`s Horizon 2020 research and innovation programme from the European Research Council (ERC) (Grant agreement No. 715063), and by the Netherlands Organisation for Scientific Research (NWO) as part of the Vidi research program BinWaves with project number 639.042.728.
SS is supported by the Australian Research Council Centre of Excellence for Gravitational Wave Discovery (OzGrav), through project number CE170100004.
AVG acknowledges support from Consejo Nacional de Ciencia y Tecnologia (CONACYT). 

Software: \\
\textsc{COMPAS}  \citep{stevenson2017formation, 2018MNRAS.477.4685B, 2018MNRAS.481.4009V}\\  \texttt{Python} available from \url{python.org}\\ matplotlib \citep{2007CSE.....9...90H}\\  \texttt{NumPy} \citep{2011CSE....13b..22V}\\ \texttt{ipython$/$jupyter} \citep{2007CSE.....9c..21P, kluyver2016jupyter} \\
This research has made use of data, software and/or web tools obtained from the Gravitational Wave Open Science Center (\url{https://www.gw-openscience.org}), a service of LIGO Laboratory, the LIGO Scientific Collaboration and the Virgo Collaboration. LIGO is funded by the U.S. National Science Foundation. Virgo is funded by the French Centre National de Recherche Scientifique (CNRS), the Italian Istituto Nazionale della Fisica Nucleare (INFN) and the Dutch Nikhef, with contributions by Polish and Hungarian institutes.




\bibliographystyle{mnras}
\bibliography{my_bib} 




\appendix

\section{Derivation of the variance used for optimising the length of the \AISs{} exploration phase}
\label{app:Fprior}
In this section we derive the expression for the variance (Eq.~\ref{eq:VarFprior}) on the rate estimate used to optimize $f_{\rm{expl}}$. 
We can estimate the optimal fraction $f_{\rm{expl}}$ of samples that we should spend in the exploratory phase by taking into account the probability of not identifying a target population forming region in the exploratory phase. We assume that we sample from the mixture distribution $Q(x)$ which is given by Eq.~\ref{eq:CombinedDistribution}:
\begin{equation}
	Q(\boldsymbol{x}) =  f_{\text{expl}}  \pi(\boldsymbol{x})  +  (1 - f_{\text{expl}}) \widetilde{q(\boldsymbol{x})}, 
\label{eq:CombinedDistributionApp}
\end{equation}
where $\pi$ is the prior (used for the exploratory phase sampling) and $\widetilde{q}$ is the mixture of Gaussians. 
We also assume we aim to estimate the rate $\rate_{\text{T}} $ with $N$ total samples.   After simulating these $N$ samples we may have identified a target binary-forming region with total weight $z_1$   whereas a region with weight $z_2$ is yet unidentified such that the estimated rate of the target population is (Eq.~\ref{eq:EstimateFexpl})%
\begin{equation}
	\rate_{\text{T}} =  \underbrace{z_1}_{\text{identified}} + \overbrace{z_2}^{\text{unidentified}} \approx {\mathbb{E}_Q[\rate_{\text{T}}]}. 	 
	\label{eq:EstimateFexplApp}
\end{equation}

The variance on $\rate_{\text{T}}$, $\mathbb{V}_Q[\rate_{\text{T}}]$, is a measure for the uncertainty of the estimated rate  ${\mathbb{E}_Q[\rate_{\text{T}}]}$. Therefore, the optimal value of  $f_{\rm{expl}}$ is one that minimizes the variance on $\rate_{\text{T}}$. To determine this we first derive an estimate for the variance  $\mathbb{V}_Q[\rate_{\text{T}}]$.

Using the continuous definition for the variance we have

\begin{equation}
	\mathbb{V}_Q[\rate_{\text{T}}] = \frac{1}{N} \left[ \int \phi(x)^2 \widetilde{w(x)}^2 Q(x) \dif x - \mathbb{E}_Q[\rate_{\text{T}}]^2  \right].
\end{equation}
where the $1/N$ factor comes from  taking the variance of the mean (average). 

Since
\begin{align}
\widetilde{w}(x) = \pi(x) / Q(x), 
\end{align}   
and  $\mathbb{E}_Q[\rate_{\text{T}}] = z_1 + z_2$, we find that

\begin{equation}
	\mathbb{V}_Q[\rate_{\text{T}}] = \frac{1}{N} \left[ \int \phi(x)^2 \widetilde{w(x)} \pi(x) \dif x - (z_1 + z_2)^2  \right].
	\label{integralB3}
\end{equation}

Since we assumed the target binary forming region is equal to $z_1 + z_2$, by definition no binaries of the target population are found outside this region and thus $\phi(x) = 0 $ outside $z_1 $ and $z_2$. Using this we can rewrite the integral in Equation \ref{integralB3} as

\begin{align}
	 \int \phi(x)^2 \widetilde{w(x)} \pi(x) \dif x  =
	  \int_{z_1}  \widetilde{w(x)} \pi(x) \dif x + \int_{z_2}  \widetilde{w(x)} \pi(x) \dif x.
	  \label{integralB4}
\end{align}

We now assume that we have found enough binaries of our target population  in our exploratory phase and that we don't have a bias for an output function such that the Gaussian mixture $\widetilde{q(x)}$ is effectively flat over $z_1$. In other words, we assume that on the target binary forming region $z_1$  
\begin{equation}
	\widetilde{q(x)}   \approx  \frac{\pi(x)}{z_1}.	 
\end{equation} 
where $\pi(x)$ is the birth distribution. 

Using this in Eq.~\ref{eq:CombinedDistributionApp}, we approximate that $Q(x) \approx   (1-f_{\rm{expl}}) \pi(x)/z_1  + f_{\rm{expl}} \pi(x)$ in $z_1$ such that
\begin{align}
	\widetilde{w(x)} \approx \frac{1}{(\frac{1-f_{\rm{expl}}}{z_1}) + f_{\rm{expl}}}  \hspace{2cm} \text{for }x \text{ in } z_1.
	\label{z1weight}
\end{align}

In addition, we also assume that the our Gaussian mixture $\widetilde{q}$ is negligible outside of the target binary forming regions, i.e. $\widetilde{q(x)} = 0 $ outside of $z_1$. In other words we assume that $z_2$ is far enough from $z_1$ that the probability is zero to sample it with $\widetilde{q}$, and that we have completely missed it during the exploratory sampling. By doing so we obtain that on $z_2$ we have
\begin{align}
	\widetilde{w(x)} \approx \frac{1}{f_{\rm{expl}}}  \hspace{2cm} \text{for }x \text{ on } z_2
	\label{z2weight}
\end{align}

Substituting  Eqs. \ref{z1weight} and \ref{z2weight} into the integral expression of Eq. \ref{integralB4}  then yields: 

\begin{align}
	 \int \phi(x)^2 \widetilde{w}(x) \pi(x)  \dif x  \approx \frac{z_1}{(\frac{1-f_{\rm{expl}}}{z_1}) + f_{\rm{expl}}} + \frac{z_2}{f_{\rm{expl}}}
	  \label{integralB5}
\end{align}
where we used $\int_{z_1} \pi(x) \dif x = z_1$ and  $\int_{z_2} \pi(x) \dif x = z_2$.

We can now write the variance as
\begin{equation}
\mathbb{V}_Q(\rate_{\text{T}}) \approx \frac{1}{N} \left[  \frac{z_1}{\frac{(1-f_{\rm{expl}})}{z_1} + f_{\rm{expl}}}  + \frac{z_2}{f_{\rm{expl}}} - (z_1 + z_2)^2    \right],	
	\label{eq:VarFpriorApp}
\end{equation} 
i.e., Eq. \ref{eq:VarFprior}.
 
\section{Toy model}
\label{app:toymodel}

\begin{figure*}
\includegraphics[width=1\textwidth]{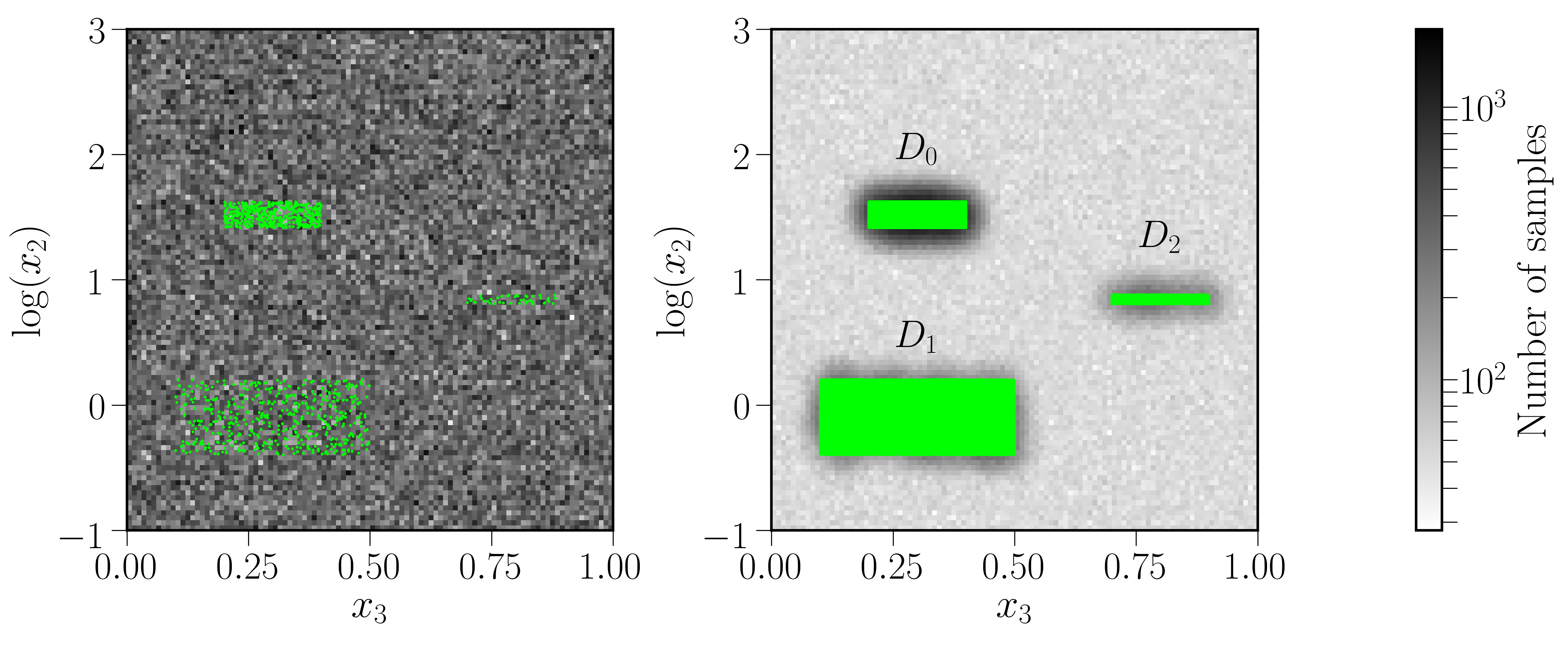}	 
\caption{Toy model illustration: distribution of the $10^6$ samples drawn in the  the birth distribution Monte Carlo (left-panel) and \AISs{}  (right panel) simulation.  The panels show two-dimensional projections of $\log x_2$ and $x_3$ from the three dimensional parameter space. In both figures the sampling density  (\textcolor{AISgray}{gray}) is shown through a two-dimensional histogram with 100$\times$100 bins. Over plotted  (\textcolor{AISgreen}{green}) are the samples that lie within the volume $V_D$ and recover the rare outcome $D$.  Dark regions surrounding the green areas of interest in the right plot indicate that our \AISs{} algorithm focuses more of the computational time around the region of interest.
\label{fig:app-3dslicedResult}
}
\end{figure*}
We construct a toy model that can be run without too much computational burden and is inspired by binary population synthesis simulations to study the performance of \AISs. The advantage of a toy model is that the moments are analytically known and the toy model can be repeatedly evaluated at minimal computational cost.  We use this toy model to investigate a suitable choice for the scale parameter $\kappa$ in the width of Gaussian sampling distributions (see Eq.~\ref{eq:sigmajk}).

We build the toy model in the 3-dimensional parameter space defined by the initial parameters $x_1, x_2$ and $x_3$. 
The distribution functions (and ranges) of $x_1, x_2$ and $x_3$ are chosen to be similar to the initial parameters $m_{\rm{1,i}}$, $a_{\rm{i}}$ and $q_{\rm{i}}$, used for our binary population synthesis model (see Appendix~\ref{app:COMPASdetails}). Similarly to the birth distribution of $a_{\rm{i}}$ in the binary population synthesis code, we sample in $\log x_2$.
The output of the toy model is constructed from a union of disconnected volumes $D$ and an output function  $\phi(\boldsymbol{x_i})$ given by 
\begin{equation}
\mathbbm{1}_{D, \rm{toymodel}}(\boldsymbol{x_i}  ) = \begin{cases} 1 \hspace{0.1cm} &\text{if } \boldsymbol{x_i} \in D \\
	0  &\text{otherwise,} 
	\end{cases}
\end{equation}
where
${D} = D_0 \cup D_1 \cup D_2$ is the union of three cuboids with the following vectors for the location of the center and the half-length of each cuboid in the $x_1, x_2$ and $ x_3$ direction 
\begin{align*}
	D_0 &= \CenterZero \pm \EpsilonZero, \notag \\
	D_1 &= \CenterOne \pm \EpsilonOne, \notag \\
	D_2 &= \CenterTwo \pm \EpsilonTwo. \notag \\
\end{align*}
The two-dimensional projected distribution of the union $D$ in $x_3$ and $\log x_2$ is shown in bright green in Figure \ref{fig:app-3dslicedResult}. One might notice that the disconnected target regions look similar to the `boats' in the game Battleship - which is often played with a strategy that is conceptually similar to \AISs{}.   The fractional rates produced by these regions are the integrals over their volumes of the local density of the birth probability distributions (the prior distribution).   The fractional rate of $D$ in the initial parameter space equals  $V_{\text{D}} = \TrueFraction$, where a prior  (birth distribution) of  $x_1 \propto x_1^{-2.3}$ is assumed on $x_1$ and flat priors are assumed on $\log x_2$, and $x_3$.  The value of $V_{\text{D}}$ is chosen to be similar to the average yield of double compact object mergers in the publicly available simulations\footnote{Populations are  available at \url{http://www.sr.bham.ac.uk/compas/data.}} of \citet{2018MNRAS.481.4009V}. In addition, the fractional rate produced from  $D_0$ is similar to that from $D_1$, whereas the fractional rate from $D_2$ is relatively 10 times smaller.  (For these parameter scalings, the absolute volume of $D_0$ is larger than the volume of $D_1$, but the different prior distributions weight the volume of $D_1$ more highly.)

We run repeated simulations varying the parameter $\kappa$ in  \AISs.  We fix the total number of samples  to $N = 10^6$.  We know the true value for the volume integral $V_D$ and calculate for each simulation the deviation between the fractional rate estimate and this true weighted volume. The closer to zero this deviation is, the better the estimate. For each variation of $\kappa$ we run 100 simulations.

The result of one such simulation for $\kappa = 2$ is shown in Fig.~\ref{fig:app-3dslicedResult}. In the \AISs{} simulation the three islands are recovered with much better resolution than in traditional Monte Carlo sampling from the prior. The dark regions around the islands $D_0, D_1 $ and $D_2$ in the \AISs{} simulation demonstrate that our method focuses more of the computational time around the regions of interest.  
In both simulations, the islands {$D_0$} and  {$D_1$} contain more samples than {$D_2$}, as expected.

\begin{figure}
	\includegraphics[width=1.1\columnwidth]{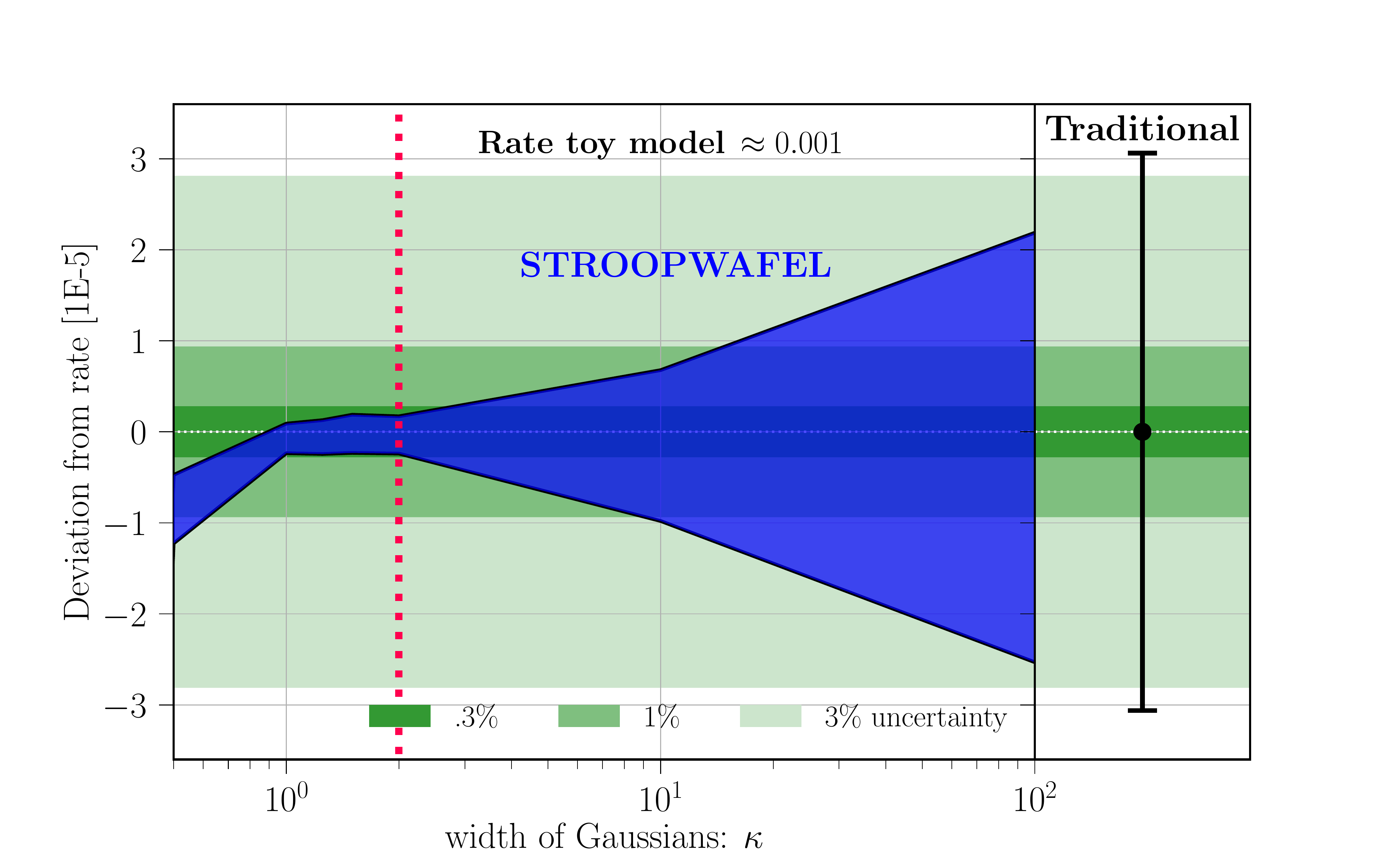}
	    \caption{Toy model test result: $1 \sigma$ deviations from the true volume integral $V_D$ when estimating the fractional rate using \AISs{} with different values for the scaling factor of the width of the Gaussians $\kappa$ (left panel).  The true fractional rate in the toy model is known and equals approximately $ 0.001$. The deviations are calculated by running 100 repeated simulations  with a total number  of $N = 10^6$ samples per simulation. 
The green contours show a $.3\%, 1\%$ and $3\%$ fractional sampling uncertainty on the rate estimate of the target population.   On the right the $1 \sigma$ uncertainty of the traditional Monte Carlo sampled simulation is shown with an error bar; this matches the expected fractional uncertainty $1/\sqrt{N_{\rm{T}}}$.  In this work we adopt $\kappa =2$ (red dotted line).} 
    \label{fig:toymodelAbsError}
\end{figure}

Figure~\ref{fig:toymodelAbsError} shows, in blue, the $1 \sigma$ deviation from the true value for \AISs \ simulations as a function of $\kappa$. The result of repeated simulations with traditional birth distribution Monte Carlo sampling is shown as a reference on the right.  Shades of green in the background show regions of 0.3$\%$, 1$\%$ and $3\%$ fractional sampling uncertainty on the rate estimate. 

Excessively small values of $\kappa$ lead to biases in the estimated rate of more than $10\%$. This is because overly narrow Gaussian distributions create ``holes'' in the adapted sampling distribution, which then no longer completely cover or characterize the regions of interest. As a result the refinement phase misses part of $D$, leading to systematic underestimation of the true rate. 

On the other hand, excessively large $\kappa$ values decrease the efficiency of the refinement phase as many samples drawn from the mixture of Gaussians $q(x)$ fall outside the target regions. The scatter from the true rate estimate approaches the scatter obtained from the traditional Monte Carlo simulation as $\kappa$ increases. 

We find that $\kappa \gtrsim 1.5$ is a robust for all our test simulations and yields substantially better estimates on rates and distribution functions compared to traditional Monte Carlo simulations. We therefore adopt $\kappa = 2$ (the red dotted line in Figure \ref{fig:toymodelAbsError}) when sampling in three dimensions.

\section{Binary population synthesis model set-up}
\label{app:COMPASdetails}

We test the performance of the algorithm \AISs \ by implementing our algorithm in the rapid binary population synthesis code \textsc{COMPAS}. 
\textsc{COMPAS} (Compact Object Mergers: Population Astrophysics and Statistics) is designed to study uncertainties in stellar binary evolution models and constrain them with observations, particularly those of gravitational-wave sources \citep{stevenson2017formation, 2018MNRAS.477.4685B, 2018MNRAS.481.4009V}.  
COMPAS interpolates between and extrapolates beyond stellar evolutionary tracks based on algorithms from the code SSE \citep{2000MNRAS.315..543H}, which rely on analytic fits of single star evolution from \citet{1998IAUS..191P.607P}. COMPAS relies on an approximate and parametrized treatment of the physical processes, including for binary interactions, and can typically compute a predicted final outcome for a single binary system within a second.  

In this work the code is used to analyse DCO systems that  form through isolated binary evolution, which often involves the common-envelope phase \citep[e.g.][]{SmarrBlandford1976}.  The approach we use to simulate a synthetic population of DCOs is similar to other binary population synthesis studies \citep[including, e.g.,][]{2002MNRAS.329..897H, belczynski2002comprehensive,Dominik:2012kk}. We evolve a population of binary systems from their birth until they form a DCO system or otherwise either merge or disrupt. We then make a sub-selection of the DCOs that consist of two compact objects that merge within the age of the Universe through gravitational-wave emission and study the properties of this population.  

In general, we follow the \Fiducial model described in \citet{2018MNRAS.481.4009V}. We mention the most important assumptions here. 
The birth distribution for the primary mass $m_{1,i}$ is chosen to be a power law distribution known as the initial mass function (IMF) where $p(m_{1,i}) \propto  m_{1,i}^{-\alpha}$ with $\alpha = 2.3$ for massive stars \citep{2001MNRAS.322..231K}. For the simulations we draw $m_{1,i} $ in $[5,150] \, \rm{M}_{\odot}$. 
The initial mass ratio $q_i = m_{2,i} / m_{1,i} $ of binary systems is suggested from observations to follow a flat distribution (e.g. \citealt{1991MNRAS.250..701T,1992ApJ...401..265M, 1994A&A...282..801G, 2012Sci...337..444S}) given by $p(q_i) \propto  1$. We adopt $q_i \in [0,1]$. 
The initial separation $a_i$ is assumed to be flat in the log, also known as \"{O}piks law  $p(a_i) \propto \frac{1 }{a_i}$ \citep{1924PTarO..25f...1O, 1983ARA&A..21..343A}. We choose $a_i \in [0.01, 1000] \rm{AU}$.   We assume that all our binaries have circular orbits at birth. 
These distributions and parameter ranges resemble commonly used settings for binary population synthesis simulations. 

Our changes to the \Fiducial model from  \citet{2018MNRAS.481.4009V}  are the following:
\begin{itemize}
\item We use a metallicity of  $Z = 0.001$  for all our simulations.  
\item We use the \texttt{DELAYED} prescription for the core-collapse supernovae treatment from \citet{2012ApJ...749...91F}. 
\item We use a prescription for pair-instability supernovae and pulsational pair instability supernovae based on \citet{Woosley_2017}.  The implementation in \textsc{COMPAS}  is described in \citet{2019arXiv190402821S}. 
\end{itemize}

We fix the total number of binaries in each simulation to $N =  10^6$ both for when using birth distribution Monte Carlo and \AISs{} sampled simulations. The total computational time for each of the birth distribution Monte Carlo simulations is approximately $180$ CPU hours. The total computational time of the \AISs{} simulations is up to a factor of 2 lower as a result of a decrease in average simulation time per sample in our sampling method compared to traditional sampling (see Section \ref{subsec:VerificationCOMPASspeedUp}). This  result rises from  binaries that become a gravitational-wave source costing on average less computational time to simulate with \textsc{COMPAS} than other binaries.

\section{Bandwidth variations of kernel density estimator}
\label{app:KDEbandwidth}

Figure~\ref{fig:mtotHistograms} and \ref{fig:Mchirpkde} use the same resolution (i.e., bandwidth or bin width) to estimate the chirp mass distribution function for the output of both the \AISs{} and birth distribution Monte Carlo simulation.  The bin width of a histogram or the bandwidth of a kernel density estimator can strongly influence the estimated distribution.  Hence, in practice, the bandwidth should be adapted to the resolution available in the data.

\begin{figure}
	\includegraphics[width=\columnwidth]{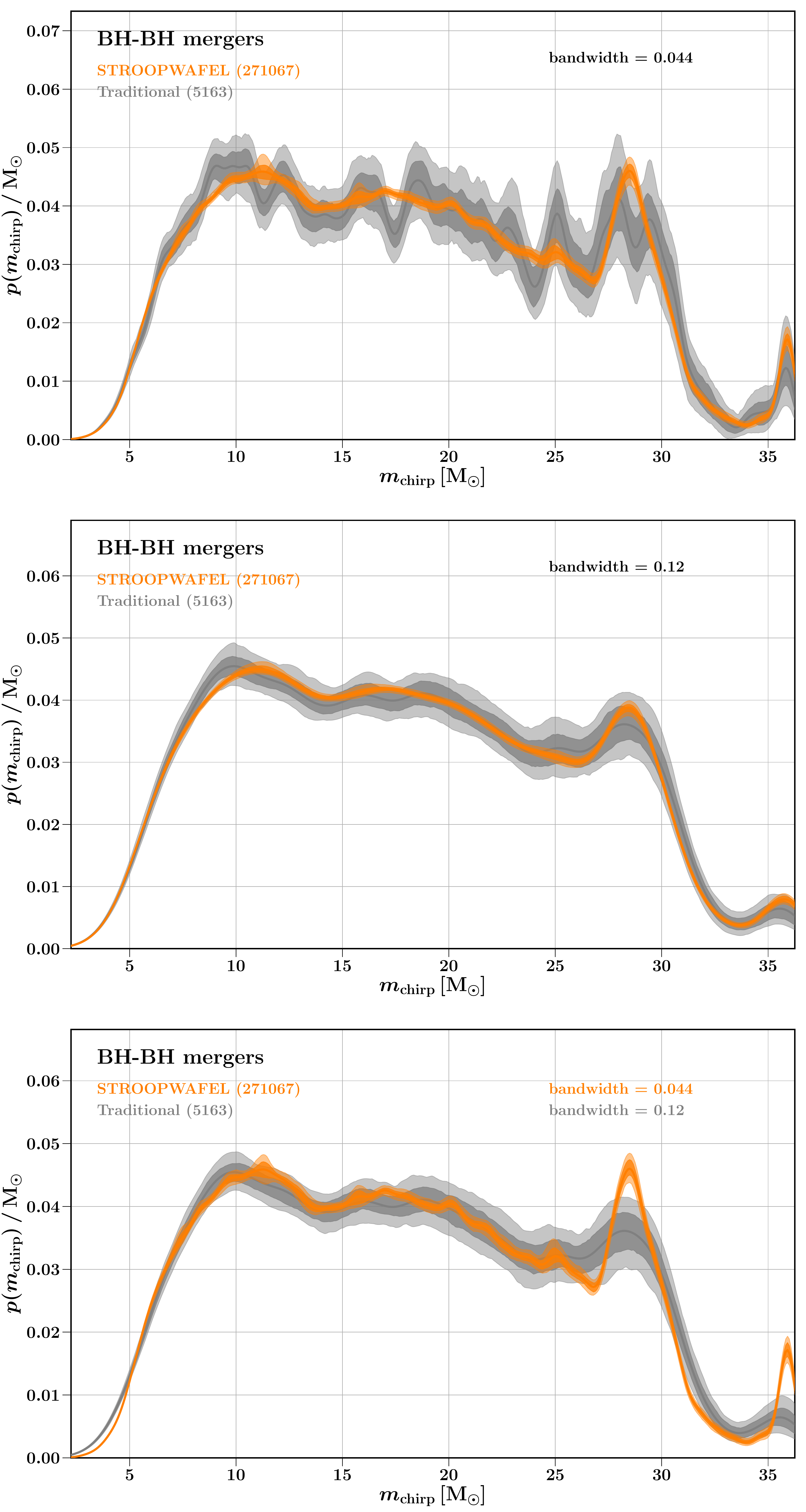}
    \caption{Predicted chirp mass distribution of the   BH--BH merger population using \AISs{} (orange) and traditional (grey) sampling. In all cases the simulation uses $N= 10^6$ samples and the distributions are weighted to the sensitivity of gravitational-wave interferometers using   \citet{2017ApJ...851L..25F}. Shaded regions show the  $1$- and $2$--$\sigma$ confidence intervals which are calculated by bootstrapping the samples 100 times. 
    This distribution is for a particular set of model assumptions, including a single metallicity $Z = 0.001$, and an integration over a metallicity-dependent cosmic star formation history is required for comparisons with observations. The same \texttt{scipy} kernel density estimator smoothing with a   kernel density estimator factor for the bandwidth of about $0.044 $ (top panel) and $0.12 $  (middle panel) is  used for the traditional and \AISs{} simulations. In the bottom panel   a  kernel bandwidth  of about $0.044 $ is used for the \AISs{} method whereas for the traditional  method we use a bandwidth of about $0.12$.  }
    \label{fig:MchirpBandwidths}
\end{figure}

We show in  Figure~\ref{fig:MchirpBandwidths} the predicted  chirp mass distribution of BH--BH  mergers using  adapted resolutions for  birth distribution Monte Carlo sampling and \AISs{} sampling.  

We estimate the adapted bandwidth using \textit{Scott`s Rule}  which, in one dimension,  scales as $\propto N_{\rm{T}}^{-1/4}$. This is the  default bandwidth choice in  the \texttt{scipy} kernel density estimator function. For the \AISs{} sampling we replace $N_{\rm{T}}$ with the effective sample size (ESS) given by $(\sum_i \widetilde{w_i})^2 / \sum_i(\widetilde{w_i}^2)$ (which, in practice, is approximately equal to $N_{\rm{T, \AISs}}$). 
The top panel of  Figure~\ref{fig:Mchirpkde} shows  the estimated  chirp mass distribution from both sampling methods for a  dimensionless kernel density estimator factor for the bandwidth of 
$\rm{ESS}_{\rm{STROOPWAFEL}}^{-1/4} \approx 0.044$. This bandwidth is too small for the $53\times$ smaller birth distribution Monte Carlo BH--BH population which therefore shows significant statistical noise  fluctuations.
The middle panel of Fig.~\ref{fig:Mchirpkde} shows  the estimated  chirp mass distribution from both sampling methods for a bandwidth  of $N_{\rm{T, traditional}}^{-1/4} \approx 0.12$.   This bandwidth causes smaller statistical fluctuations for the traditional sampling, but removes some of the more detailed features for the \AISs{} sampled distribution.  In the bottom panel the two plots are combined, showing the distributions with the relative bandwidths. The \AISs{} obtains smaller uncertainties on the distribution as well as a higher resolution, which is a result from the higher number of BH--BH mergers found in this simulation.

%


\bsp	
\label{lastpage}
\end{document}